\documentclass[aps,prc,twocolumn,superscriptaddress,showpacs,preprintnumbers,amsmath,amssymb]{revtex4}
\usepackage{graphicx}
\usepackage{dcolumn}
\usepackage{bm}
\usepackage{colordvi}
\usepackage{array}
\usepackage{booktabs}

\def\eqlab#1{\label{eq:#1}}
\def\eqref#1{Eq.~(\ref{eq:#1})}

\begin{document}

\def\CAT{INFN-Sezione di Catania, I-95123 Catania, Italy}
\def\ULIV{University of Liverpool, Physics Department, Liverpool L69 7ZE, United Kingdom}
\def\UNAM{Instituto de F\'{i}sica, Universidad Nacional Aut\'{o}noma de M\'{e}xico, A.P. 20-364, M\'{e}xico 01000 D.F., Mexico}
\def\BUCH{IFIN-HH, Reactorului 30, 077125 M\v{a}gurele-Bucharest, Romania}
\def\GSI{GSI Helmholtzzentrum f\"{u}r Schwerionenforschung GmbH, D-64291 Darmstadt, Germany}
\def\ZAGREB{Ru{d\llap{\raise 1.22ex\hbox
   {\vrule height 0.09ex width 0.2em}}\rlap{\raise 1.22ex\hbox
   {\vrule height 0.09ex width 0.06em}}}er
   Bo\v{s}kovi\'{c} Institute, HR-10002 Zagreb, Croatia}
\def\TUD{Technische Universit\"at Darmstadt, D-64289 Darmstadt,  Germany}
\def\JAGU{M. Smoluchowski Institute of Physics, Jagiellonian University, Pl-30-348 Krak\'ow, Poland}
\def\STFC{STFC Daresbury Laboratory, Warrington WA4 4AD, United Kingdom}
\def\HUZHOU{School of Science, Huzhou University, Huzhou 313000, P.R. China}
\def\IFJ{H. Niewodnicza{\'n}ski Institute of Nuclear Physics, Pl-31342 Krak{\'o}w, Poland}
\def\CENBG{CENBGn Universit\'e de Bordeaux, CNRS/IN2P3, F-33175 Gradignan, France}
\def\GANIL{GANIL, CEA et IN2P3-CNRS, F-14076 Caen, France}
\def\LNS{INFN-Laboratori Nazionali del Sud, I-95123 Catania, Italy}
\def\KACST{KACST, Riyadh, Saudi Arabia}
\def\SAUD{Physics Department, King Saud University, Riyadh,  Saudi Arabia}
\def\MESSINA{INFN-Gruppo Collegato di Messina, I-98166 Messina, Italy}
\def\UMESS{Dipartimento di Scienze Matematiche e Informatiche, Scienze Fisiche e Scienze della Terra, University of Messina, I-98166 Messina, Italy}
\def\SANT{Universidade de Santiago de Compostela, 15782 Santiago de Compostela, Spain}
\def\MILAN{INFN-Sezione di Milano, I-20133 Milano, Italy}
\def\TEXAS{Department of Chemistry and Cyclotron Institute, Texas A$\&$M University, College Station, TX-77843, USA}
\def\MSU{Department of Physics and Astronomy and NSCL, Michigan State University, East Lansing, MI-48824, USA}
\def\WMICH{Western Michigan University, Kalamazoo, MI-49008, USA}
\def\UCAT{Dipartimento di Fisica e Astronomia-Universit\`a, I-95123 Catania, Italy}
\def\POLIMILANO{Dipartimento di Elettronica, Informazione e Bioingegneria, Politecnico di Milano, I-20133 Milano, Italy}
\def\UMILANO{Dipartimento di Fisica, Universit\`a degli Studi di Milano, I-20133 Milano, Italy}
\def\TORINO{INFN and DISAT, Politecnico di Torino, I-10129 Torino, Italy}
\def\RIKEN{RIKEN, Wako, Saitama 351-0198, Japan}
\def\DEBR{Institute for Nuclear Research (MTA Atomki), P.O. Box 51, H-4001 Debrecen, Hungary}
\def\ENNA{Universit\`{a} degli Studi di Enna "Kore", I-94100 Enna, Italy}
\def\NAPOLI{INFN-Sezione di Napoli, I-80126 Napoli, Italy}
\def\UNAPOLI{Dipartimento di Fisica "Ettore Pancini", Universit\`{a} di Napoli Federico II, I-80126 Napoli, Italy}
\def\BRATIS{Institute of Physics, Slovak Academy of Sciences, 84511 Bratislava 45, Slovakia}
\def\LMU{Fakult\"{a}t f\"{u}r Physik, Universit\"{a}t M\"{u}nchen, D-85748 Garching, Germany}
\def\WU{Chemistry Department, Washington University, St. Louis, MO-63130, USA}
\def\IPNO{Institut de Physique Nucl\'{e}aire, IN2P3-CNRS et Universit{\'e} Paris-Sud, F-91406 Orsay, France}

\title{Results of the ASY-EOS experiment at GSI: The symmetry energy \\at suprasaturation density}

\affiliation{\CAT}
\affiliation{\ULIV}
\affiliation{\JAGU}
\affiliation{\LNS} 
\affiliation{\UNAM}
\affiliation{\KACST} 
\affiliation{\SAUD} 
\affiliation{\MESSINA} 
\affiliation{\UMESS} 
\affiliation{\TUD}
\affiliation{\GSI}
\affiliation{\SANT} 
\affiliation{\ZAGREB}
\affiliation{\GANIL}
\affiliation{\IFJ}
\affiliation{\TEXAS} 
\affiliation{\MSU}
\affiliation{\WMICH}
\affiliation{\BUCH}
\affiliation{\UCAT}
\affiliation{\MILAN}
\affiliation{\POLIMILANO}
\affiliation{\UMILANO}
\affiliation{\TORINO} 
\affiliation{\RIKEN} 
\affiliation{\DEBR} 
\affiliation{\ENNA} 
\affiliation{\STFC}
\affiliation{\HUZHOU}
\affiliation{\NAPOLI} 
\affiliation{\UNAPOLI} 
\affiliation{\CENBG}
\affiliation{\WU} 
\affiliation{\IPNO} 
\affiliation{\BRATIS} 
\affiliation{\LMU} 

\author{P.~Russotto}            \affiliation{\CAT}
\author{S.~Gannon}              \affiliation{\ULIV}
\author{S.~Kupny}               \affiliation{\JAGU}
\author{P.~Lasko}               \affiliation{\JAGU}
\author{L.~Acosta}              \affiliation{\LNS}\affiliation{\UNAM}
\author{M.~Adamczyk}            \affiliation{\JAGU}
\author{A.~Al-Ajlan}            \affiliation{\KACST}
\author{M.~Al-Garawi}           \affiliation{\SAUD}
\author{S.~Al-Homaidhi}         \affiliation{\KACST}
\author{F.~Amorini}             \affiliation{\LNS}
\author{L.~Auditore}            \affiliation{\MESSINA}\affiliation{\UMESS}
\author{T.~Aumann}              \affiliation{\TUD}\affiliation{\GSI} 
\author{Y.~Ayyad}               \affiliation{\SANT}
\author{Z.~Basrak}              \affiliation{\ZAGREB}
\author{J.~Benlliure}           \affiliation{\SANT}
\author{M.~Boisjoli}            \affiliation{\GANIL}
\author{K.~Boretzky}            \affiliation{\GSI}
\author{J.~Brzychczyk}          \affiliation{\JAGU}   
\author{A.~Budzanowski}\thanks{deceased}         \affiliation{\IFJ}
\author{C.~Caesar}              \affiliation{\TUD}
\author{G.~Cardella}            \affiliation{\CAT}
\author{P.~Cammarata}           \affiliation{\TEXAS}
\author{Z.~Chajecki}            \affiliation{\MSU}\affiliation{\WMICH}
\author{M.~Chartier}            \affiliation{\ULIV}
\author{A.~Chbihi}              \affiliation{\GANIL}
\author{M.~Colonna}             \affiliation{\LNS}
\author{M.~D.~Cozma}            \affiliation{\BUCH}
\author{B.~Czech}               \affiliation{\IFJ}
\author{E.~De~Filippo}          \affiliation{\CAT}
\author{M.~Di~Toro}             \affiliation{\LNS}\affiliation{\UCAT}
\author{M.~Famiano}             \affiliation{\WMICH}
\author{I.~Ga\v{s}pari\'c}      \affiliation{\TUD}\affiliation{\ZAGREB}
\author{L.~Grassi}              \affiliation{\ZAGREB}
\author{C.~Guazzoni}            \affiliation{\MILAN}\affiliation{\POLIMILANO}
\author{P.~Guazzoni}            \affiliation{\MILAN}\affiliation{\UMILANO}
\author{M.~Heil}                \affiliation{\GSI}
\author{L.~Heilborn}            \affiliation{\TEXAS}
\author{R.~Introzzi}            \affiliation{\TORINO}
\author{T.~Isobe}               \affiliation{\RIKEN}
\author{K.~Kezzar}              \affiliation{\SAUD}
\author{M.~Ki\v{s}}             \affiliation{\GSI}
\author{A.~Krasznahorkay}        \affiliation{\DEBR}
\author{N.~Kurz}                \affiliation{\GSI}
\author{E.~La~Guidara}           \affiliation{\CAT}
\author{G.~Lanzalone}           \affiliation{\LNS}\affiliation{\ENNA}
\author{A.~Le~F\`evre}          \affiliation{\GSI}
\author{Y.~Leifels}             \affiliation{\GSI}
\author{R.~C.~Lemmon}           \affiliation{\STFC}
\author{Q.~F.~Li}               \affiliation{\HUZHOU}
\author{I.~Lombardo}            \affiliation{\NAPOLI}\affiliation{\UNAPOLI}
\author{J.~{\L}ukasik}          \affiliation{\IFJ}
\author{W.~G.~Lynch}            \affiliation{\MSU}
\author{P.~Marini}              \affiliation{\GANIL}\affiliation{\TEXAS}\affiliation{\CENBG}
\author{Z.~Matthews}            \affiliation{\ULIV}
\author{L.~May}                 \affiliation{\TEXAS}
\author{T.~Minniti}             \affiliation{\CAT}
\author{M.~Mostazo}             \affiliation{\SANT}
\author{A.~Pagano}              \affiliation{\CAT}
\author{E.~V.~Pagano}           \affiliation{\LNS}\affiliation{\UCAT}
\author{M.~Papa}                \affiliation{\CAT}
\author{P.~Paw{\l}owski}        \affiliation{\IFJ}
\author{S.~Pirrone}             \affiliation{\CAT}
\author{G.~Politi}              \affiliation{\CAT}\affiliation{\UCAT}
\author{F.~Porto}               \affiliation{\LNS}\affiliation{\UCAT}
\author{W.~Reviol}              \affiliation{\WU}
\author{F.~Riccio}              \affiliation{\MILAN}\affiliation{\POLIMILANO}
\author{F.~Rizzo}               \affiliation{\LNS}\affiliation{\UCAT}
\author{E.~Rosato}\thanks{deceased}              \affiliation{\NAPOLI}\affiliation{\UNAPOLI}
\author{D.~Rossi}               \affiliation{\TUD}\affiliation{\GSI}
\author{S.~Santoro}             \affiliation{\MESSINA}\affiliation{\UMESS}
\author{D.~G.~Sarantites}                \affiliation{\WU}
\author{H.~Simon}               \affiliation{\GSI}
\author{I.~Skwirczynska}         \affiliation{\IFJ}
\author{Z.~Sosin}\thanks{deceased}                 \affiliation{\JAGU}
\author{L.~Stuhl}               \affiliation{\DEBR}
\author{W.~Trautmann}           \affiliation{\GSI}
\author{A.~Trifir\`o}           \affiliation{\MESSINA}\affiliation{\UMESS}
\author{M.~Trimarchi}           \affiliation{\MESSINA}\affiliation{\UMESS}
\author{M.~B.~Tsang}             \affiliation{\MSU}
\author{G.~Verde}               \affiliation{\CAT}\affiliation{\IPNO}
\author{M.~Veselsky}            \affiliation{\BRATIS}
\author{M.~Vigilante}           \affiliation{\NAPOLI}\affiliation{\UNAPOLI}
\author{Yongjia~Wang}                \affiliation{\HUZHOU}
\author{A.~Wieloch}             \affiliation{\JAGU}
\author{P.~Wigg}                \affiliation{\ULIV}
\author{J.~Winkelbauer}         \affiliation{\MSU}
\author{H.~H.~Wolter}           \affiliation{\LMU}
\author{P.~Wu}                 \affiliation{\ULIV}
\author{S.~Yennello}            \affiliation{\TEXAS}
\author{P.~Zambon}              \affiliation{\MILAN}\affiliation{\POLIMILANO}
\author{L.~Zetta}               \affiliation{\MILAN}\affiliation{\UMILANO}
\author{M.~Zoric}               \affiliation{\ZAGREB}


\date{\today}

\begin{abstract}
Directed and elliptic flows of neutrons and light charged particles were measured for the reaction 
$^{197}$Au+$^{197}$Au at 400 MeV/nucleon incident energy within the ASY-EOS experimental campaign at the 
GSI laboratory. The detection system consisted of the Large Area Neutron Detector LAND, combined with 
parts of the CHIMERA multidetector, of the ALADIN Time-of-flight Wall, and of the Washington-University 
Microball detector. The latter three arrays were used for the event characterization and reaction-plane 
reconstruction. In addition, an array of triple telescopes, KRATTA, was used for complementary 
measurements of the isotopic composition and flows of light charged particles.

From the comparison of the elliptic flow ratio of neutrons with respect to charged particles with UrQMD predictions, 
a value $\gamma = 0.72 \pm 0.19$ is obtained for the power-law coefficient describing the density dependence
of the potential part in the parametrization of the symmetry energy. 
It represents a new and more stringent constraint for the regime of suprasaturation density and confirms, 
with a considerably smaller uncertainty, the moderately soft to linear density dependence deduced from 
the earlier FOPI-LAND data. The densities probed are shown to reach beyond twice saturation.

\pacs{21.65.Cd, 21.65.Ef, 25.75.Ld}

\end{abstract}
\maketitle

\section{Introduction}
\label{sec:int}

Differences in the collective emission properties of neutrons and protons in neutron-rich heavy-ion reactions
at intermediate bombarding energies have been proposed as potential observables for the study of the 
equation of state of asymmetric nuclear matter~\cite{baoan96,baoan02,greco03,yong06}.
Among them, the neutron-proton elliptic-flow ratio and difference have been shown to be sufficiently
sensitive probes of the high-density behavior of the nuclear symmetry energy~\cite{Rus11,Coz11}. 
The comparison of existing data from the FOPI-LAND experiment~\cite{Lei93,lamb94} 
with calculations performed with the UrQMD transport model~\cite{qli05,qli06,Li:2006ez} suggested a 
moderately soft to linear symmetry term, characterized by a coefficient $\gamma = 0.9 \pm 0.4$ 
for the power-law parametrization of the density dependence of the potential part
of the symmetry energy~\cite{Rus11}. 
This result has excluded super-soft scenarios but suffers from the considerable statistical uncertainty 
of the experimental data.
 
The same data set was also compared to calculations performed with 
the QMD model originally developed in T\"{u}bingen~\cite{khoa92,maheswari98} 
and a constraint compatible with the UrQMD result was obtained~\cite{Coz11,Coz13,Rus14}. In addition, 
a thorough study of the parameter dependence of the model predictions was performed 
to devise a route towards a model-independent constraint of the high-density symmetry 
energy. It showed that presently acceptable limits for the choice of parameters in the 
isoscalar part of the transport description cause uncertainties comparable with but not larger 
than those of the experimental FOPI-LAND data~\cite{Coz13}.
It was also found that different parametrizations of the 
isovector part of the equation of state, the Gogny-inspired (momentum-dependent, Ref.~\cite{Das:2002fr})  
and the power-law (momentum-independent) potential, lead to very similar predictions for the 
neutron-vs-charged-particle elliptic-flow ratio or difference. 

To improve the statistical accuracy of the experimental flow parameters for 
the $^{197}$Au+$^{197}$Au reaction and to extend the flow measurements to other systems, 
the symmetric collision 
systems $^{197}$Au+$^{197}$Au, $^{96}$Zr+$^{96}$Zr, and $^{96}$Ru+$^{96}$Ru at 400 MeV/nucleon 
incident energies have been chosen for the asymmetric-matter equation-of-state (ASY-EOS) experimental campaign, conducted at the GSI laboratory 
in May 2011 (experiment S394). As in the FOPI-LAND experiment, the Large Area Neutron Detector (LAND)~\cite{LAND}  
was used for the detection and identification of neutrons and light charged 
particles. Parts of the CHIMERA multidetector~\cite{Pag04,DeFilippo14}, of the ALADIN Time-of-flight 
Wall~\cite{schuettauf96}, and of the Washington University Microball detector~\cite{BALL} 
were used for the event characterization and determination of the azimuthal reaction-plane orientation.   
By including the Krak\'{o}w Triple Telescope Array (KRATTA)~\cite{Luk11} with isotopic identification 
of charged-particles up to atomic number $Z=4$ in the setup, additional observables as, e.g., 
yields and flows of light-charged particles and yield ratios of the isobar pairs $^{3}$H/$^{3}$He 
or $^{7}$Li/$^{7}$Be were made available for the study of 
isospin effects in these reactions.

The results reported here refer exclusively to the $^{197}$Au+$^{197}$Au reaction whose
analysis has been completed. It is shown that the new data confirm the moderately soft to 
linear density dependence of the symmetry energy deduced from the earlier FOPI-LAND data. 
However, for technical reasons, 
the capabilities of the LAND detector could not be fully exploited. This had the effect that
the originally intended measurement of detailed dependencies of the neutron flows on rapidity, transverse
momentum, and particle type could not be fully realized. Uncertainties of some of the required
corrections  restricted the analysis to essentially only providing the ratio of neutron over 
charged-particle flows, integrated over the LAND acceptance. By comparing it with the 
results of UrQMD calculations adapted to the experimental acceptance and analysis conditions, 
a new and more stringent constraint for the symmetry energy at suprasaturation densities was derived. 

The technical deficiencies of the LAND timing system, the methods developed to correct for them in the analysis, 
and the consequences for the obtained results are described and explained in detail in the Appendix. 
The confidence in the validity of the main, acceptance integrated, result is derived from the fact that it 
is found to be only weakly dependent on assumptions regarding details of the corrections. These 
uncertainties were quantitatively assessed by varying the assumptions within well-defined intervals and by 
treating their effects as systematic errors. These systematic and the statistical errors of the
collected data set are of approximately equal magnitude.

The present work derives its importance also from the fact that the flow probe, at present, appears to
be the most robust observable for testing the nuclear equation of state at high densities. 
The recent comprehensive study of charged-particle flows for $^{197}$Au+$^{197}$Au collisions at 
energies from 0.4 to 1.5 GeV/nucleon reflects a remarkable consistency in its support of a soft solution
for the equation of state of symmetric matter, including momentum-dependent forces~\cite{reisdorf12,lefevre15}. 
It provides a narrower constraint than previously available~\cite{dani02}. 
Such narrower limits for the compressibility 
of symmetric nuclear matter are very useful also with regard to the equation of state of asymmetric matter.
They have the effect of reducing systematic uncertainties originating from the choice of parameters for 
the isoscalar sector of a transport description~\cite{Coz13}. 

Major efforts have recently been made to reduce the apparent systematic discrepancies in the interpretation of 
the FOPI pion ratios~\cite{reis07} with increasingly complex transport 
calculations~\cite{hong14,song15,yong16,baoanli15,wenmeiguo15,cozma16}. 
Of particular interest is the observation that the predicted $\pi^-/\pi^+$ yield ratios are expected
to rise when the medium modifications of pion production thresholds are explicitly considered~\cite{song15}. 
This effect may permit reproducing the experimental values with choices for the symmetry energy that are
less extreme than those required in some of the earlier pion studies~\cite{xiao09,feng10,xie13}. 
However, the 
calculations of Hong and Danielewicz~\cite{hong14} exhibit only a small sensitivity of integrated pion ratios
to the stiffness of the symmetry energy, pointing to the need for energy-differential observables.
Further work will thus be required before pion yields and yield ratios 
can be reliably applied to the investigation of the high-density symmetry energy.

The important role played by the nuclear symmetry energy in nuclear structure and reactions as well as in 
astrophysics is the subject of several review articles~\cite{baran05,lattprak07,lipr08,ditoro10,gandolfi15}.
A brief introductory review of the situation at suprasaturation densities 
is available in Ref.~\cite{Traut12}. A comprehensive list of pertinent articles has recently appeared
in Topical Issue on Nuclear Symmetry Energy~\cite{EPJA2014}.

\section{Experimental details}
\label{sec:exp}

\subsection{Setup for S394}
\label{sec:setup}

\begin{figure}[htb]            
\centering
\includegraphics[width=8.5cm]{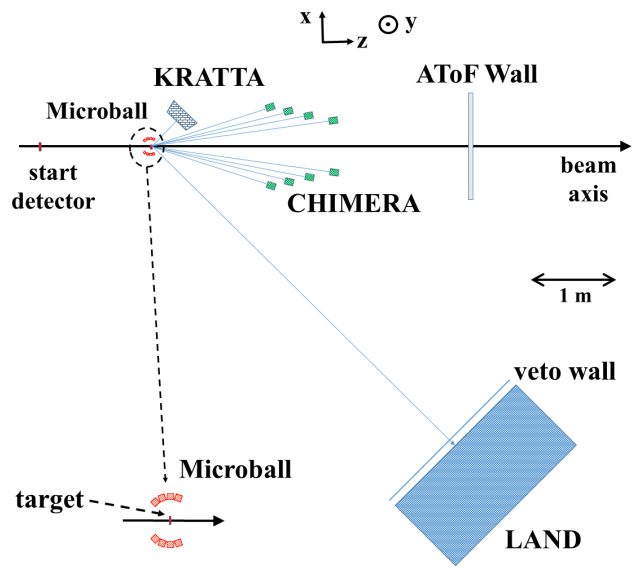}
\caption{Schematic view of the experimental setup of the ASY-EOS experiment S394 at GSI. The chosen coordinate system is indicated;
the $y$ direction points upwards in the laboratory. The target area with the
Microball is not to scale in the main drawing but shown with a scale factor of approximately 5:1 in the lower left corner
(see Sec.~\protect\ref{sec:det} for coverage and dimensions).
}
\label{fig:setup}       
\end{figure}

A schematic view of the experimental setup of the ASY-EOS experiment at the GSI laboratory
is shown in Fig.~\ref{fig:setup}. The beam was
guided in vacuum to about 2~m upstream from the target. A thin plastic scintillator foil viewed by two
photomultipliers was used to record the projectile arrival times and to serve as a start 
detector for the time-of-flight measurement.
The Large Area Neutron Detector, LAND \cite{LAND}, was positioned to cover laboratory angles 
around 45$^{\circ}$ with respect to the beam direction. A veto wall of plastic 
scintillators in front of LAND allowed discriminating between neutrons and charged particles. 
In this configuration, it was possible to measure the directed and elliptic flows of neutrons and 
charged particles near midrapidity within the same angular acceptance.
Opposite of LAND, covering a comparable range of polar angles, the Krak\'{o}w Triple Telescope 
Array, KRATTA~\cite{Luk11}, had been installed to permit flow measurements of identified 
charged particles under the same experimental conditions. Results obtained with KRATTA will 
be published separately.

For the event characterization and for measuring the orientation of the reaction plane, three
detection systems had been installed. The ALADIN Time-of-Flight (AToF) Wall~\cite{schuettauf96} was used 
to detect charged particles and fragments in the forward direction at polar angles up to
$\theta_{\rm lab} \le 7^{\circ}$. Its capability of identifying large fragments and of characterizing
events with a measurement of $Z_{\rm bound}$~\cite{schuettauf96} permitted the sorting of events 
according to impact parameter. Four double rings of the CHIMERA multidetector~\cite{Pag04,DeFilippo14} 
carrying together 352 CsI(Tl) scintillators in the forward direction and four rings with 50 thin CsI(Tl) 
elements of the Washington University Microball array~\cite{BALL} surrounding the target
provided sufficient coverage and granularity for determining the orientation 
of the reaction plane from the measured azimuthal particle distributions. 

\begin{figure}[ht]       
\centering
\includegraphics[width=8.5cm]{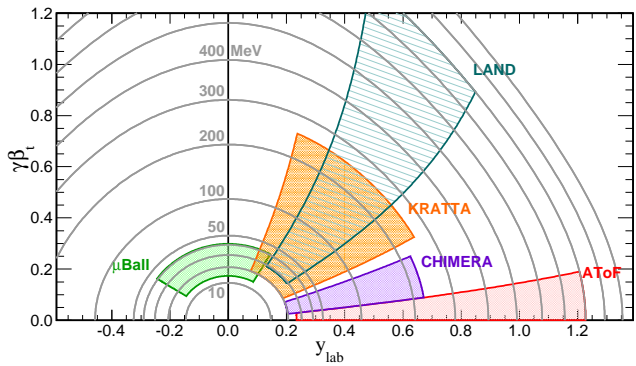}
\caption{Kinematic acceptance in the transverse-velocity vs rapidity plane of the detector systems used
in the S394 experiment. The contour lines refer to the specified values of the kinetic energy of protons
in the laboratory, ranging from 10 MeV to 1 GeV.
The indicated lower and upper limits in energy are for protons (stopped protons for
KRATTA and CHIMERA) and were calculated for the specific detector thresholds and configurations. 
An average value was chosen for the four types of detector elements of the Microball (labeled
$\mu$Ball in the figure).
}
\label{fig:accept}       
\end{figure}

\begin{figure}[htb]       
\centering
\includegraphics[width=8.5cm]{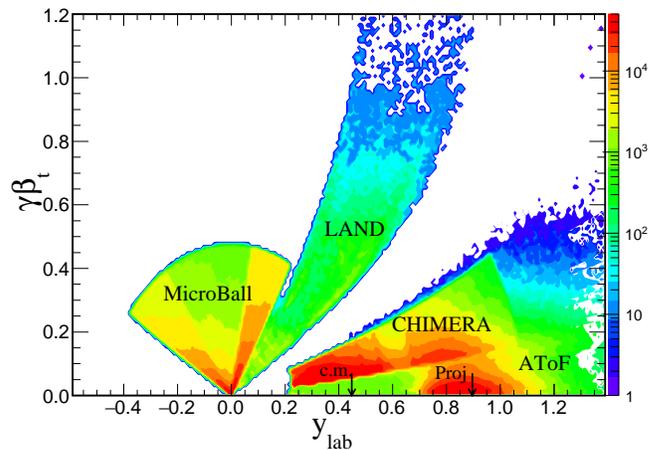}
\caption{Measured invariant hit distribution for $^{197}$Au+$^{197}$Au collisions at 
400 MeV/nucleon incident energy in the transverse-velocity vs rapidity plane for 
charged particles detected with the three systems Microball, CHIMERA, and AToF Wall with
full azimuthal coverage and for neutrons detected with LAND. 
The velocities of particles detected with the Microball are not measured and shown
here with an arbitrarily chosen homogeneous kinetic-energy distribution in the interval
$0 \le E_{\rm kin} \le 100 MeV$. The apparent angular variation may be influenced by
ring-dependent thresholds.
The arrows indicate the rapidities of the projectile $y_p = 0.896$ 
and of the c.m. system.
}
\label{fig:inv}       
\end{figure}

The kinematic coverage
achieved with this assembly of detection systems is illustrated in 
Figs.~\ref{fig:accept} and~\ref{fig:inv}. In Fig.~\ref{fig:inv}, in particular, the enhanced particle
yields in the kinematic regimes of participant and spectator emissions are clearly visible. The
product yields from the decay of the projectile spectator seen with CHIMERA and the AToF Wall do not 
exactly match because the AToF efficiency for hydrogen isotopes in this energy range is lower than that
of the CHIMERA modules.

\subsection{Detection systems}
\label{sec:det}

\subsubsection{LAND detector}
\label{sec:land}

The Large Area Neutron Detector, LAND~\cite{LAND}, upgraded 
with new TACQUILA electronics developed at GSI~\cite{Koch05}, was positioned at a distance of 5~m from the target. 
Its kinematic acceptance was similar to that of the forward LAND subdetector used in the 
FOPI-LAND experiment~\cite{Rus11} but slightly larger in rapidity for given transverse momentum 
owing to the shorter distance from the target. LAND consists of 10 consecutive layers of $2 \times 2$~ m$^2$ area, 
together adding up to the 1-m depth of the detector. Each layer is formed by 20 modules 
of 2-m length whose orientations alternate from layer to layer between vertical and horizontal. 
The modules have a $10 \times 10$~cm$^2$ cross section and are built from nine sheets of iron and ten sheets of
plastic-scintillator material, all 5 mm thick, arranged in alternating order and oriented parallel 
to the entrance plane of the detector. 
Two iron sheets of 2.5 mm thickness form the entrance and exit layers of each module. In this design,
the iron serves as a converter and the plastic scintillators as detectors for the produced ionizing radiation.

As it turned out during the analysis, the standard method of identifying the 
showers generated by interacting neutrons in the full LAND assembly was not feasible because of the timing
difficulties related to the use of the new electronic system (discussed in Sec.~\ref{sec:landtiming} 
below and in the Appendix). Only 19 modules (out of 20) of the first 
layer of LAND are included in the present analysis. This lowers the detection efficiency for neutrons 
and modifies its energy dependence, effects that had to be taken into account. The resulting 
range of polar angles that were covered by this part of LAND was 
$37.7^{\circ} \le \theta_{\rm lab} \le 56.5^{\circ}$ with respect to the beam direction. 

A veto wall consisting of 10-cm-wide and 5-mm-thick plastic-scintillator slabs 
covered the front face of LAND, permitting the distinction between neutral and charged particles.
The slabs were mounted in vertical orientation parallel to the modules of the first plane of LAND.
Charged particles were identified on the basis of coincident hits in the veto wall, matching the 
time and position of the corresponding hit in LAND. However, 
owing to insufficient resolution achieved in the readout of the analog signals, the identification
of the atomic number $Z$ of the recorded charged particles on the basis of their energy loss in the 
veto-wall scintillators was not feasible. The comparative analysis was thus restricted to the 
collective flows of neutrons with respect to that observed for all charged particles detected
within the acceptance of LAND.

\subsubsection{KRATTA hodoscope}
\label{sec:kratta}

The Krak\'{o}w Triple Telescope Array, KRATTA \cite{Luk11}, was specifically designed for the 
experiment to measure the energy, emission angles, and isotopic composition of light charged reaction 
products. The 35 modules of KRATTA were arranged in a $7 \times 5$~array and placed opposite to LAND at a 
distance of 40 cm from the target. They covered 160 msr of solid angle at polar angles 
between 24$^{\circ}$ and 68$^{\circ}$.
The modules of KRATTA consisted of two, optically decoupled, CsI(Tl)
crystals (thickness of 2.5 and 12.5~cm) and three large-area, 500-$\mu$m-thick, PIN photodiodes.
The middle photodiode and the short CsI(Tl) crystal read out by the diode from its front face were operated as
a single-chip telescope~\cite{pasquali91}.
Very good isotopic resolution has been obtained in the whole dynamic range up to Z $\sim$ 4. The
methods used for deriving it and the virtue of using digital pulse-shape recording throughout are 
described in Ref.~\cite{Luk11}.

\subsubsection{CHIMERA  hodoscope}
\label{sec:chimera}

Four double rings of the CHIMERA multidetector~\cite{Pag04,DeFilippo14} had been transported to the GSI laboratory 
and installed at their nominal distances from the target, covering polar angles between 7$^{\circ}$ and 20$^{\circ}$. 
They carried together 352 CsI(Tl) scintillators, 12~cm in thickness and read out with photodiodes.
Each of the eight individual rings provided
a $2\pi$ azimuthal coverage with either 40 or 48 modules per ring. For calibration purposes, four of the 
Si detectors of the regular CHIMERA setup were installed in each ring. For these telescopes,
an independent digital pulse-shape acquisition system was used to investigate and improve the particle 
identification and calibration methods~\cite{Ac13}. The recorded telescope data proved 
very useful for verifying the analysis schemes developed for this experiment.

The CHIMERA rings were intended for the detection and identification of light-charged particles, primarily
expected to come from the midrapidity regime. In the analysis, a rapidity gate $y > 0.1$ in the 
center-of-mass (c.m.) reference system was applied to exclusively select forward-hemisphere 
emissions for determining the orientation of the reaction plane.

\begin{figure}[htb]              
\centering
\includegraphics[width=8.5cm]{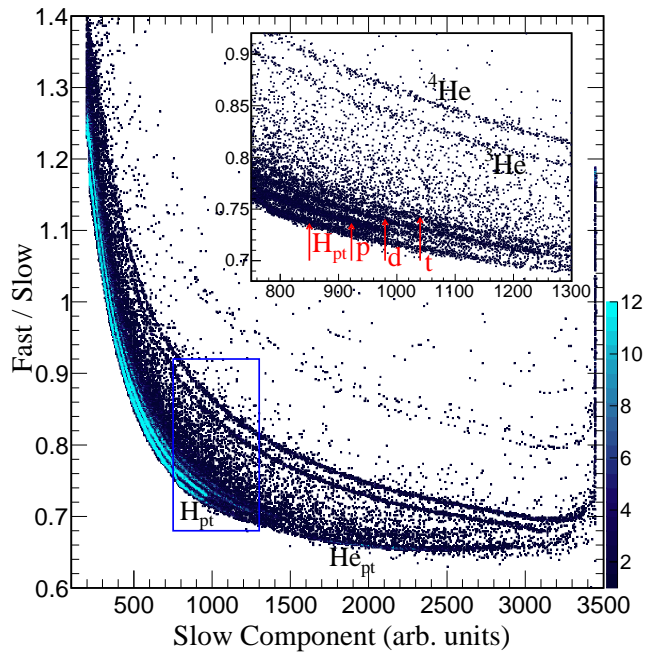}
\caption{Identification plot of CsI(Tl) signals recorded with a CHIMERA module of ring 7 
($\theta_{\rm lab} \approx 17^{\circ}$) from $^{197}$Au+$^{197}$Au collisions at 400 MeV/nucleon
displaying the ratio of fast-over-slow vs the slow signal components. The loci of
hydrogen and helium ions punching through the full length of the detector are labeled as H$_{\rm pt}$ and He$_{\rm pt}$.
An expanded view of the area within the rectangular box is shown in the inset. 
Besides the punch-through groups, also the loci of mass-identified light ions are indicated there.
}
\label{fig:chim1}       
\end{figure}

For the use of CHIMERA modules at the present energy regime, the identification of punch-through 
particles was essential. In addition, the velocity of registered particles had to be reconstructed with an 
accuracy permitting the application of the rapidity gate. 
For particles stopped in the CsI, this was done using the mass number $A$ and the deposited energy of the 
particles resolved in the fast-vs-slow identification map. 

For particles punching through the CsI, their atomic number, essentially $Z = 1$ or 2, was evident in the 
fast-vs-slow identification plots. 
A most probable mass number $A$ was assigned on the basis of the measured energy loss $\Delta E$ and 
used to reconstruct the total kinetic energy 
and momentum. The mass $A=4$ was assigned to helium isotopes.
In the case of the hydrogen isotopes, $A=3$ was assigned to a $Z=1$ particle 
if $\Delta E_{d}^{p.t.} < \Delta E < \Delta E_{t}^{p.t.}$, $A=2$ was assigned if 
$\Delta E_{p}^{p.t.} < \Delta E < \Delta E_{d}^{p.t.}$, 
and $A=1$ was assigned if $\Delta E < \Delta E_{p}^{p.t.}$. Here $\Delta E_{x}^{p.t.}$ refers to the 
calculated maximum energy loss $\Delta E$ deposited in the CsI(Tl) module by punch-through particles,  
and the subscript $x = p, d, t$ indicates protons, deuterons and tritons, respectively.
The reconstructed total kinetic energy was then used to determine the velocity of the particle.
An example of the two-dimensional maps used for the 
particle identification and analysis is shown in Fig.~\ref{fig:chim1}.

\subsubsection{ALADIN ToF Wall}
\label{sec:tof}

A central square part of the ALADIN Time-of-Flight (AToF) Wall \cite{schuettauf96} with an area of
approximately 1 m$^2$ was placed symmetrically with respect to the beam direction at a distance of 3.7~m downstream from the target. It
was used to detect forward emitted charged particles and fragments at polar angles smaller than 7$^{\circ}$, i.e. within
the opening of the forward-most CHIMERA ring. The two layers of the AToF Wall (front and rear) each consisted 
of 48 modules of $2.5 \times 110$~cm$^{2}$ plastic scintillators with a thickness of 1~cm and 
with photomultipliers mounted at their upper and lower end faces. The modules are arranged in densely 
packed groups of eight modules, six groups per layer, and all oriented in vertical direction. They provided
the atomic numbers $Z$ of the detected fragments and light charged particles, as well as their velocities 
and directions of emission. The threshold was set below the maximum of the $Z=1$ distribution in the spectrum of 
recorded energy-loss signals. A central hole of $7.5 \times 7.5$~cm$^2$ permitted the non-interacting beam 
to pass undetected through the AToF Wall. 

\begin{figure}[!t]              
\centering
\includegraphics[width=8.5cm]{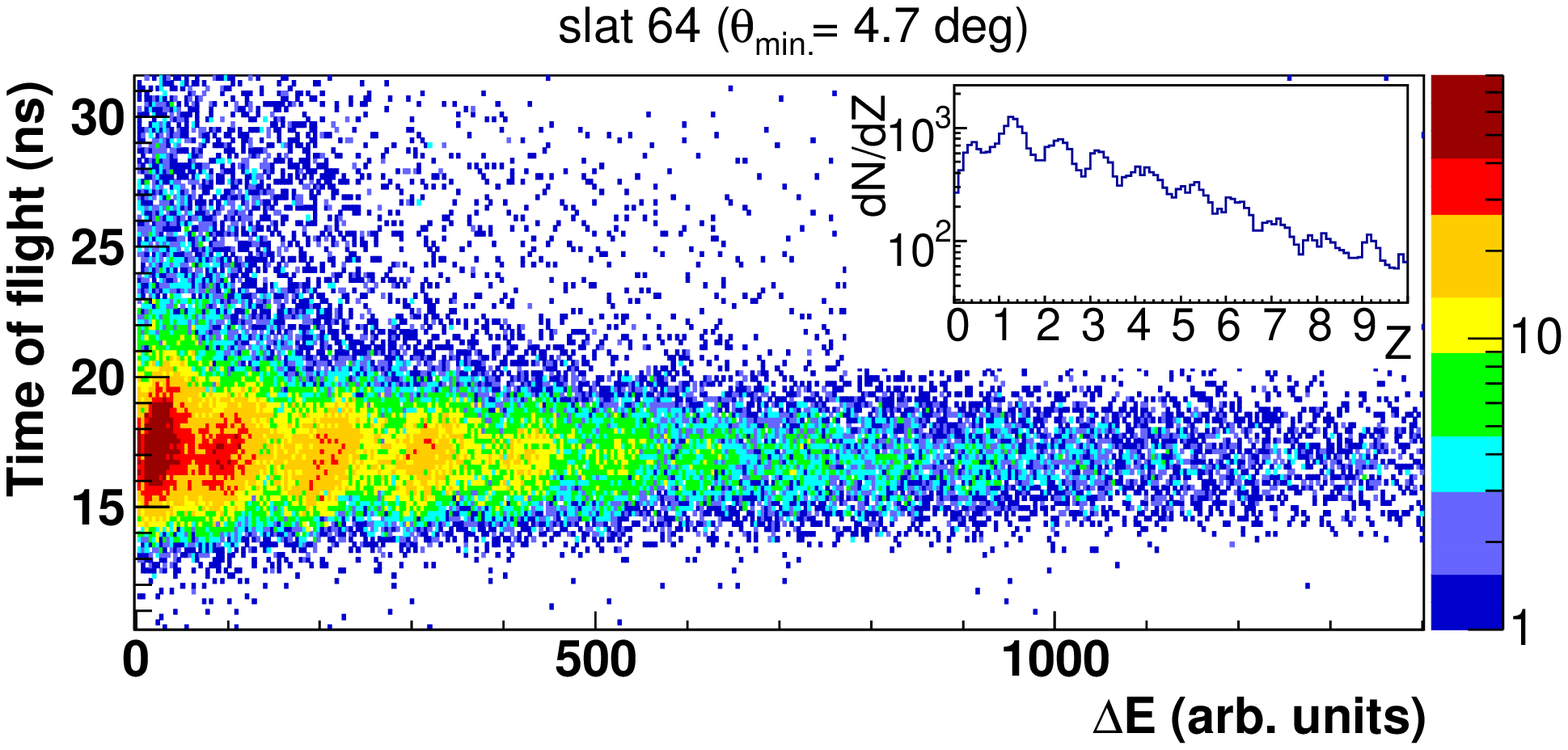}
\vspace{0.5cm}
\includegraphics[width=8.5cm]{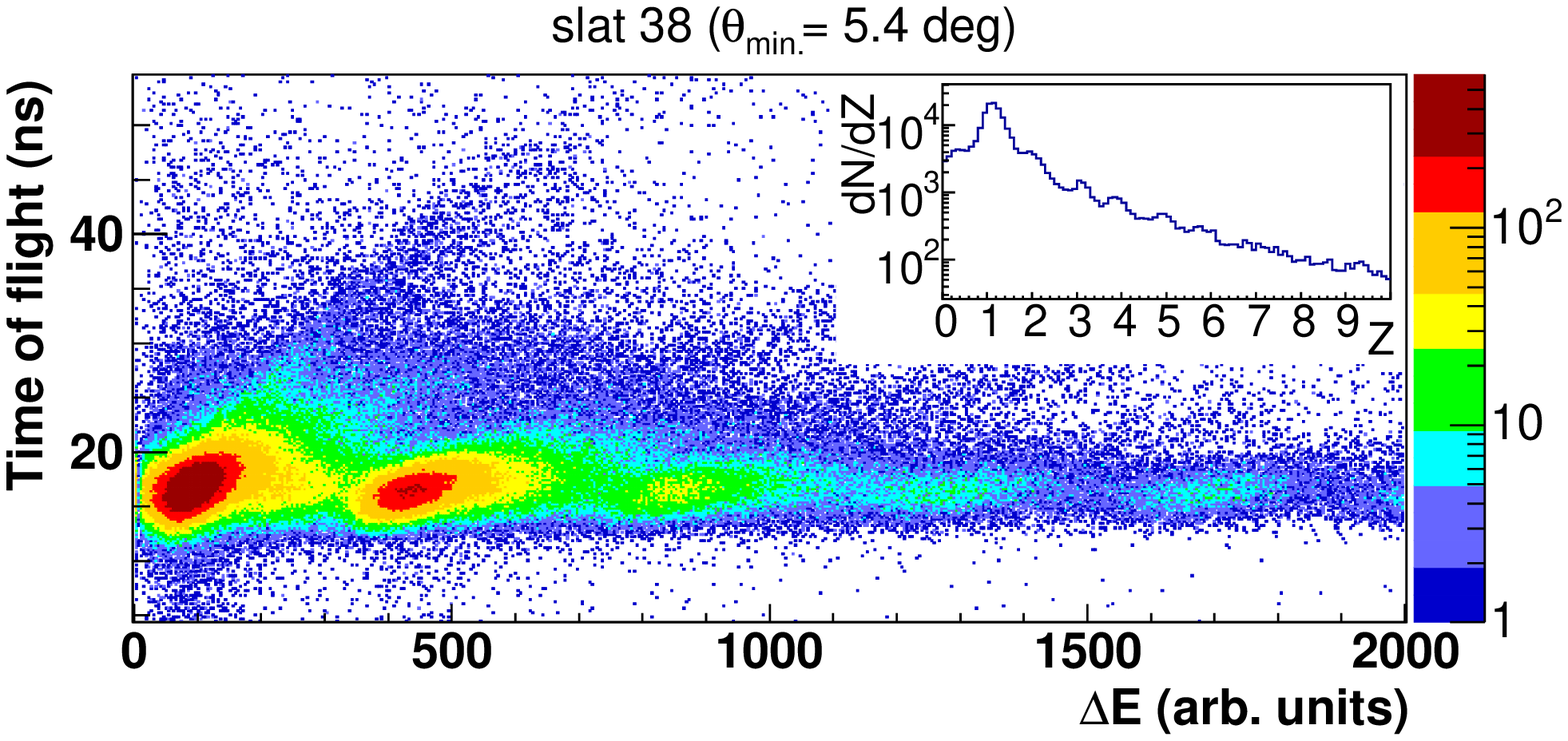}
\vspace{-0.6cm}
\caption{AToF identification plots of calibrated time of flight vs recorded energy loss $\Delta E$ for
two slats approximately 30 and 35 cm to the right of the beam direction ($\theta_{\rm lab} \approx 4.7^{\circ}$ and
$\approx 5.4^{\circ}$, respectively, at their central parts). The groups of light elements 
are clearly recognized up to atomic number $Z \approx 10$ as shown in the insets on logarithmic scales.
}
\label{fig:atof}       
\end{figure}

The atomic number $Z$ of light fragments is individually resolved on the basis of the measured time and 
energy loss up to approximately $Z=10$, as illustrated in Fig.~\ref{fig:atof}. The unusually high background
appearing in these maps is attributed to interactions of the ions with air during their flight path to
the detector. 
Heavier fragments are 
identified with a resolution of $\Delta Z \approx 2$ (FWHM) on the basis of the $Z$ calibrations generated 
in earlier experiments with the AToF Wall~\cite{schuettauf96,ogul11}.
The 
time-of-flight resolution varies with $Z$, smoothly decreasing from 300 ps (standard deviation) 
for lithium fragments to about 100 ps for fragments with $Z > 10$. The AToF timing signals were used 
to generate a reaction trigger. The minimum requirement was three recorded tracks in the front-wall modules and 
three recorded tracks in the rear-wall modules. The front- and rear-wall tracks are usually pairwise correlated 
and produced by the same particles. The central group of eight modules containing the central opening was not 
included in the trigger circuit. These trigger requirements had the effect of suppressing collisions 
producing moderate excitations. However, the forward position of the wall and the long passage of the beam 
through air had the effect of still producing 
unwanted trigger signals generated by reactions on non-target material. The methods chosen to efficiently 
eliminate such events in the analysis are explained below.

\subsubsection{Microball}
\label{sec:micro}

The target was surrounded by an array of 50 3.6-to-5.6-mm-thick CsI(Tl) elements of the Washington University 
Microball (so-called Reaction Microball~\cite{BALL}). This array had four azimuthally symmetric rings, 
subtended the range of polar angles between 60$^{\circ}$ and 147$^{\circ}$ in the 
laboratory, and thus was essentially sensitive to backward emissions in the c.m. frame of the reaction. The
azimuthal distributions of modules recording a hit above threshold provided a measure of the orientation of the 
reaction plane as seen in the rear hemisphere. 
The small diameter of the array of only about 10 cm offered a nearly negligible solid angle 
for reactions occurring downstream from the target, a property that was used for suppressing 
background reactions in the analysis. 

\subsection{Beams and targets}
\label{sec:beam}

With beam intensities of about $10^{5}$ pps and targets of 1-2\% interaction probability, 
about $5 \times 10^{6}$ events were collected for each of the systems $^{197}$Au+$^{197}$Au, 
$^{96}$Zr+$^{96}$Zr, and $^{96}$Ru+$^{96}$Ru. Additional runs were performed  
without a target to measure the background from the interaction of projectile 
ions with non-target material. The 3.7~m column of air between the target and the AToF Wall represents by itself an additional target with a 
theoretical interaction probability of about 6\% for $^{197}$Au projectiles. 

Measurements with iron shadow bars in front of LAND, with and without a target, were used 
to determine the background of scattered neutrons not directly originating from the target. 
The shadow bars consisted of several pieces of iron,
together representing a block of 60 cm in thickness and shaped to precisely cover the solid-angle
acceptance of the LAND detector as seen from the target position. Results obtained with 
the $^{96}$Zr and $^{96}$Ru beams and targets are not presented here.

\section{Data analysis}
\label{sec:datana}

The analysis of the experimental data has been performed within the FairRoot software framework 
primarily developed for the use with the future Facility for Antiproton and Ion Research (FAIR) detectors~\cite{fairroot}.
The FairRoot framework contains a complete simulation of the ASY-EOS detector setup and geometry 
and of the data analysis schemes. Theoretical calculations can be performed within the same software 
environment and filtered to adapt them to the experimental acceptance and analysis conditions. 

\subsection{LAND timing}
\label{sec:landtiming}

A major difficulty arose from the fact that the new TACQUILA electronics~\cite{Koch05} 
of the LAND detector did not permit the recognition of the very-low-energy $\gamma$-ray signals
in the LAND modules. The absolute time calibration, therefore, had to be obtained from a spectra comparison
with data of the FOPI-LAND experiment. Furthermore, the digital timing information was found to be frequently, 
with approximately 30\% -40\% probability, affected by $\pm 25$~ns time jumps, arising from errors 
in counting the number of 25-ns clock cycles occurring between the start and the stop signals in a time 
measurement. These uncertainties were identified and corrected with procedures that are
described in detail in the Appendix. Where possible, recourse was taken by comparing with or adjusting 
to existing data from previous FOPI and FOPI-LAND experiments.

\begin{figure}[!t]            
\includegraphics[width=7cm]{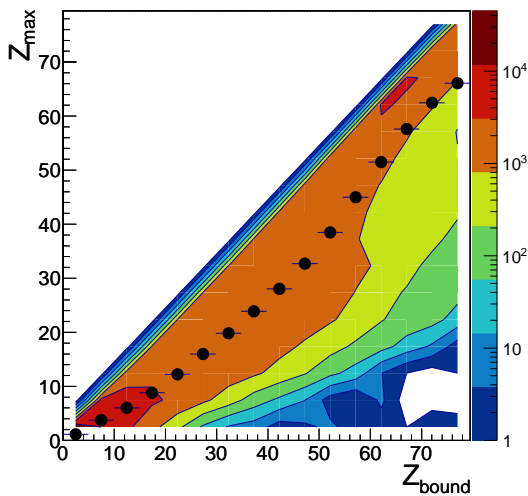}
\caption{Experimental correlation of the maximum atomic number, $Z_{\rm max}$, of the fragments 
within an event and the quantity $Z_{\rm bound}$ as deduced from the fragments 
detected with the AToF Wall. The dots represent the mean values of $Z_{\rm max}$ over the intervals of
$Z_{\rm bound}$ indicated by the horizontal error bars.
}
\label{fig:corrzmax} 
\end{figure}

The goal pursued in the present analysis consisted in applying the evident corrections 
and in quantifying the uncertainties associated with correction steps that could 
not be unambiguously determined. For the time-resolved differential data, the main uncertainty arises
from the so-called second correction step, devised for wrongly recorded hits
not recognized in the first correction step (see the Appendix). In addition to
recovering the correct times of the intended class of hits, it has the side effect of
misplacing an unknown number of valid hits in the time spectra. This causes a mixing of the flow 
properties within the affected time intervals. The problem was investigated by applying the second 
correction to randomly chosen fractions of the selected group of candidate hits and by comparing the 
consequences with data sets obtained in FOPI measurements~\cite{reisdorf2013}. 
It is shown that the mixing affects the deduced flow parameters but,
to a much smaller extent, the flow ratios. Its contribution to the
systematic error of the power-law exponent $\gamma$ amounts to $\Delta \gamma = \pm 0.05$.  

\begin{figure}[!t]            
\includegraphics[width=7cm]{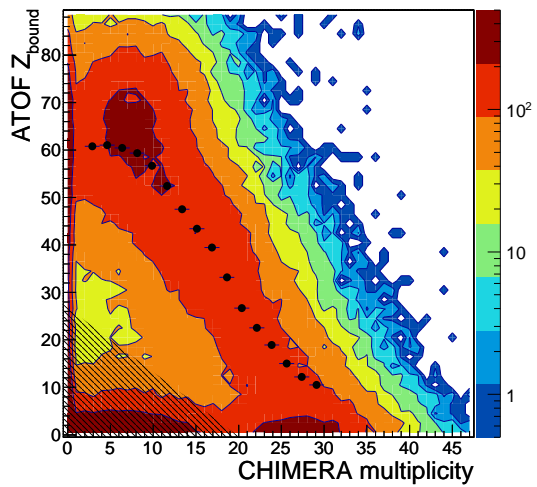}
\caption{Experimental correlation of the quantity $Z_{\rm bound}$ as deduced from the 
fragments detected with the AToF Wall with the charged-particle multiplicity measured with CHIMERA.
Events within the hatched area were excluded from the analysis; the high-intensity group near multiplicity 27 with 
$Z_{\rm bound} \approx 0$ is caused by central collisions;
the symbols represent the mean 
$Z_{\rm bound}$ of the remaining distribution as a function of the CHIMERA multiplicity.
}
\label{fig:corrmult} 
\end{figure}

This particular correction and the mixing that it causes play only a minor role for the acceptance-integrated results 
obtained after integrating over the full time spectra. Timing errors have no consequence here as long as they do
not lead beyond the limits of the integration interval. A remaining source of uncertainty is the precise 
choice of the low-energy thresholds as it should match their counterparts in the calculations. 
For charged particles, the threshold energy is given by the requirement to pass through the
veto wall and to reach the first scintillator plane of LAND, for protons about 60 MeV. It is thus independent of the
time measurement, provided the hit is within the accepted time interval.
For neutrons, the low-energy threshold is defined by the chosen integration limit at long times-of-flight.
Timing errors are effective here. To minimize the overall uncertainty, the integration limit was placed at times of flight
much longer than expected for charged particles and into a low-intensity region less affected by the
timing corrections (see the Appendix). Its nominal value corresponded to 30 MeV kinetic energy for nucleons.
The level of remaining uncertainties was determined by varying the integration limit within a wide interval
and by comparing with calculations performed with corresponding energy thresholds for neutrons. 
As observed in the differential case, the flow ratios are only mildly affected because uncertainties cancel.
The observed variation of $\Delta \gamma = \pm 0.07$ represents the overall systematic error arising
from the LAND timing properties.

\subsection{Impact parameter determination}
\label{sec:impact}

For selecting according to impact parameter, global variables were constructed from the 
CHIMERA and AToF data. 
They included 
\begin{equation}
Z_{\rm bound}=\sum_{i=1}^{N} Z_{i}~{\rm with}~Z_{i}\geq2 
\end{equation}
and the ratio of transverse to longitudinal charge,
\begin{equation} 
{\rm ZRAT}=10 Z_{\rm trans}/Z_{\rm long} 
\end{equation}
with an arbitrarily chosen scale factor 10 and with
\begin{equation} 
Z_{\rm trans}=\sum_{i=1}^{N} Z_{i}sin^{2}(\theta_{i}),~
Z_{\rm long}=\sum_{i=1}^{N} Z_{i}cos^{2}(\theta_{i})
\end{equation} 
where $\theta_{i}$ is the polar angle of the $i^{th}$ particle in the laboratory reference system. $Z_{\rm bound}$ 
is close to the charge of the primary spectator system and 
monotonically correlated with the impact parameter, while ZRAT increases with the centrality of the reaction.
The choice of these variables as impact parameter selectors has been guided by performing UrQMD 
calculations for given impact parameter ranges and filtering the simulated reaction events 
for angular acceptance, detection thresholds, and resolution of the detectors. 

For constructing $Z_{\rm bound}$, fragments recorded with CHIMERA and the AToF Wall were used where 
not otherwise indicated.
Larger fragments ($Z > 4$) are exclusively expected at very forward angles, well within the kinematic acceptance 
of $\theta_{\rm lab} \le 7^{\circ}$ of the 
AToF Wall (cf. Figs.~\ref{fig:chim1} and \ref{fig:atof}). 
The evolution of the largest atomic number, 
$Z_{\rm max}$, 
observed in an event as a 
function of $Z_{\rm bound}$, here from AToF alone, is shown in Fig.~\ref{fig:corrzmax}. 
The relative behavior of these two observables 
resembles closely that known from earlier results reported by the ALADIN Collaboration for the $^{197}$Au+$^{197}$Au 
reaction~\cite{schuettauf96,traut07}. Only for large $Z_{\rm bound}$ is a difference observed, 
as $\langle Z_{\rm max} \rangle$ 
does not reach up as close to the projectile $Z$ as it did in the ALADIN experiments with different trigger conditions. 
The trigger chosen for the present experiment suppressed the most peripheral events with a small 
multiplicity of charged particles and a corresponding $Z_{\rm max}$ near $Z=79$. 

The expected anticorrelation of $Z_{\rm bound}$ as determined from AToF alone, 
rising with impact parameter $b$ and the multiplicity of charged particles measured with the CHIMERA rings at
intermediate angles is observed as well (Fig.~\ref{fig:corrmult}). The group of events with both small
$Z_{\rm bound}$ and small multiplicities detected with CHIMERA (hatched area in the figure) is
interpreted as containing nearly undeflected heavy projectile fragments that have
passed undetected through the central hole of the AToF Wall. Such events are expected from very peripheral 
$^{197}$Au+$^{197}$Au collisions as well as from the interaction of the beam with N or
O nuclei of the air downstream of the target. The class of events within the hatched region was not further
considered in the analysis.

\begin{figure}[!t]            
\centering
\includegraphics[width=8.5cm]{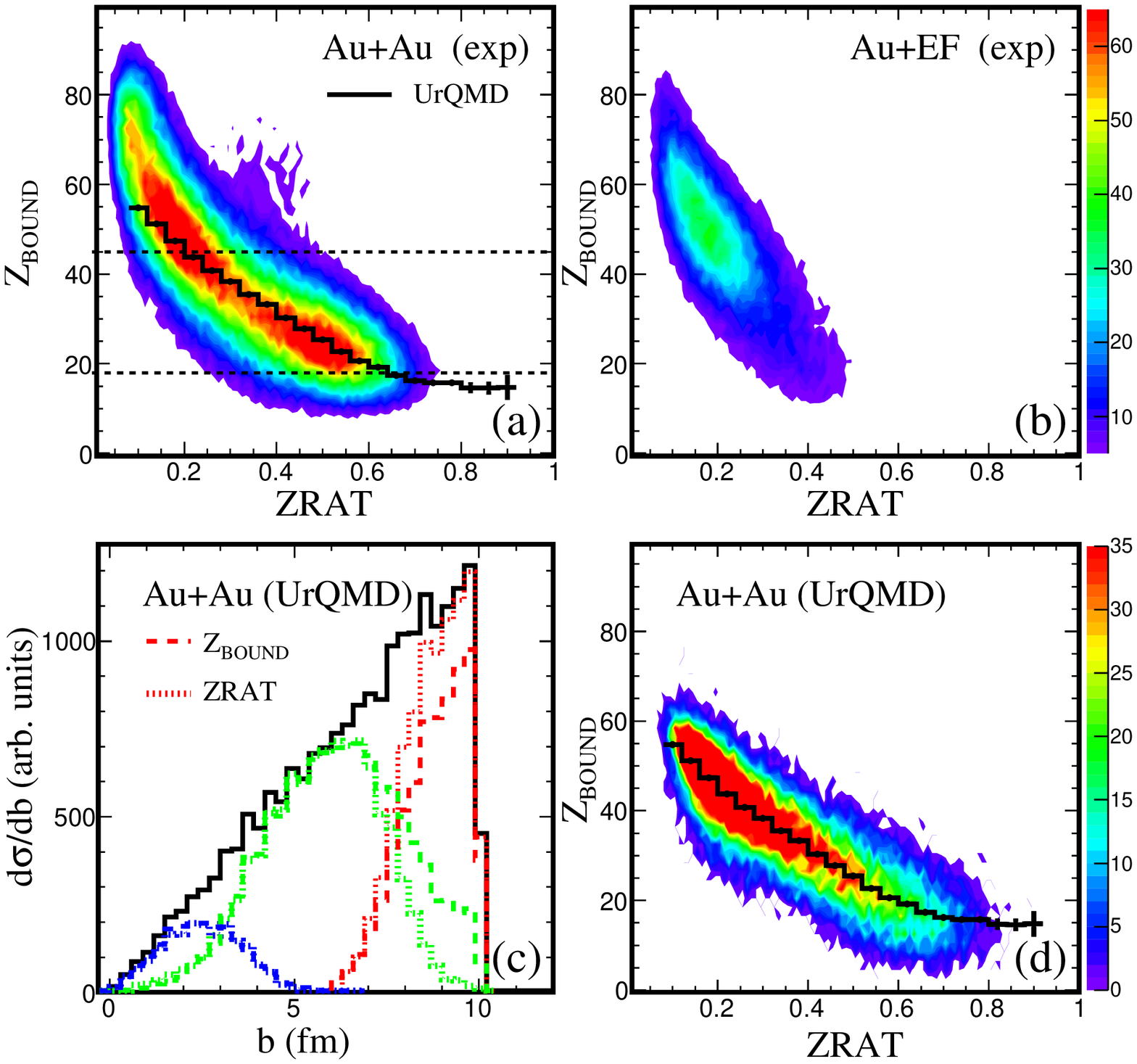}
\caption{(Top row) Inclusive $Z_{\rm bound}$ vs 
ZRAT correlation for data sets taken with (a) and without (b) a target foil in place 
(EF stands for empty frame). \\
(Bottom row) UrQMD calculations for the correlation of $Z_{\rm bound}$ vs ZRAT 
for $^{197}$Au+$^{197}$Au collisions at 400 MeV/nucleon and impact parameter $b<10.0$~fm, filtered to 
match the experimental conditions (d), and for impact-parameter distributions d$\sigma$/d$b$ obtained under 
various conditions (c). The unbiased distribution for the full reaction for $b<10.0$~fm is
given by the black (solid) histogram while the blue, green, and red lines show impact parameter distributions 
obtained when selecting  
very central, semicentral, and peripheral event classes, respectively, by gating either on
$Z_{\rm bound}$ (dashed) or on ZRAT (dotted, see Table~\protect\ref{tab:impact}).
The dashed horizontal lines in (a) represent the corresponding gates for the $Z_{\rm bound}$ selection.
The line of $Z_{\rm bound}$ centroids as a function of ZRAT of the UrQMD distribution of (d) is drawn into 
the experimental distribution (a).
}
\label{fig:impact}       
\end{figure}

The correlation of $Z_{\rm bound}$ with ZRAT, as obtained from the combined CHIMERA and AToF data 
for $^{197}$Au+$^{197}$Au collisions at 400 MeV/nucleon, are presented in Fig.~\ref{fig:impact} (a). 
The two impact-parameter sensitive quantities are globally 
anticorrelated as expected: $Z_{\rm bound}$ grows while ZRAT drops with 
increasing impact parameter. For orientation, ZRAT = 0.15 is obtained for particles detected 
at the forward limit
of the CHIMERA acceptance $\theta_{\rm lab} = 7^{\circ}$, ZRAT = 1.3 for particles detected at the largest
angle $\theta_{\rm lab} = 20^{\circ}$, and ZRAT~$\approx 0.7$ for a homogeneous distribution within the 
CHIMERA acceptance. The observed distribution is compatible with these limits. Values smaller than 
ZRAT = 0.15 are suppressed by the trigger condition of four or more charged particles 
detected with CHIMERA and two or more hits recorded by the Microball by which peripheral collisions are suppressed. 
In addition, the adopted condition requiring an anticorrelation of the preferential azimuthal directions 
of these particles observed with CHIMERA and with the Microball was applied (see Sec.~\ref{sec:corr}). 
A very similar pattern is observed for the result
of UrQMD calculations, performed for the range of impact parameters $b<10.0$~fm [Fig.~\ref{fig:impact}(d)].  
The centroid line deduced from the simulations follows
the experimental distribution shown in panel (a) rather well.

The correlation observed when the target foil is removed is shown in panel (b) 
of Fig.~\ref{fig:impact}. The yields are normalized with respect to the integrated beam 
intensity, so that the much lower intensity of background reactions becomes evident. They display a 
similar anticorrelation, however much less pronounced and extending mainly over the range typical for the 
more peripheral collisions in the $^{197}$Au+$^{197}$Au case. The observed concentration of background
events at large $Z_{\rm bound} > 40$ also coincides with the expectation for collisions of $^{197}$Au beam
particles with predominantly $^{14}$N encountered downstream of the target location~\cite{hubele91}. The
initially high yield of AToF trigger signals from $^{197}$Au+air collisions is reduced to the apparent 
low level by 
applying the conditions on the multiplicity and azimuthal orientation of Microball hits within the event. 

\begin{table}[h]
\begin{tabular}{|l|c|c|c|c|c|c|c|}
\hline
\toprule                   
Class, $b$ interval        & \multicolumn{3}{c}{$Z_{\rm bound}$} \vline & \multicolumn{3}{c}{ZRAT} \vline \\ \hline
\midrule
                           & min & max & $<b>$ & min & max & $<b>$ \\ \hline
very central, $<3.0$ fm    &  0 & 18 & 2.56 & 0.615 & 2.0   & 2.51  \\ \hline
semi-c, $3.0-7.5$ fm       & 18 & 45 & 6.18 & 0.245 & 0.615 & 5.71 \\ \hline
peripheral, $>7.5$ fm      & 45 &    & 8.74 &       & 0.245 & 8.76 \\ \hline
central, $<7.5$ fm         & 0  & 45 & 5.69 & 0.245 & 2.0   & 5.27 \\ \hline
FOPI, $3.35-6.0$ fm        & 19 & 33 & 5.00 & 0.365 & 0.585 & 4.69 \\ \hline
\end{tabular}
\caption{Selection gates used to define the indicated five classes of centrality. Their names
and the nominal ranges of impact parameter $b$ are given in the first column (semi-c stands for semicentral).
The gate required for the comparison with FOPI data (Sec.~\protect\ref{sec:timcorr}) is given in the bottom row. 
The following columns list the
minimum and maximum values of the gating intervals used and the corresponding mean values of the impact parameter $b$ as given by
the UrQMD calculations for the two sorting variables $Z_{\rm bound}$ (columns 2-4) 
and ZRAT (columns 5-7). No upper gate of $Z_{\rm bound}$
and no lower gate of ZRAT was applied when selecting peripheral events. 
}
\label{tab:impact}
\end{table}

For the actual impact-parameter selections within the range of interest $b<7.5$~fm, the global observables 
$Z_{\rm bound}$ and ZRAT were used. 
The intervals chosen to select very central, semicentral and peripheral event classes are listed in Table~\ref{tab:impact}
together with the mean impact parameters expected for these classes according to the UrQMD calculations.
The condition on multiplicity specified above provided no additional
restriction within this range of central and semicentral collisions [cf. Fig.~\ref{fig:impact}(a)]. The quality of the resolution
that can be expected, according to the UrQMD model, is illustrated in panel (c) of 
Fig.~\ref{fig:impact}. The examples of very 
central, semicentral and peripheral selections with nominal impact-parameter intervals of
$b<3$~fm, $3<b<7.5$~fm, and $b>7.5$~fm, respectively, are displayed. The expected smoothing of the
boundaries of the actually selected intervals is about equal for the $Z_{\rm bound}$ and ZRAT
observables. The interval chosen for generating the acceptance-integrated flow ratio in the final
analysis is a nominal $b<7.5$~fm, listed as central class in the table. As the calculations show, the actual distribution can be expected 
to contain nearly all events with $b<6$~fm and, with decreasing probability, a selection of events with
impact parameters up to $b \approx 10$~fm.

\subsection{Reaction plane orientation}
\label{sec:reacplane}

For the experimental estimates of the azimuthal orientation of the impact-parameter vector, both CHIMERA 
and AToF data were used. In the CHIMERA analysis, a $Q$ vector~\cite{Dan85} was calculated as 
\begin{equation}
\vec{Q}_{\rm CHI}= \sum_{i=1}^{N} Z_{i}\vec{\beta_{t,i}} \gamma_{i}, 
\label{eq:qvec}
\end{equation}
with the transverse-velocity vectors $\vec{\beta_{t,i}}$ and with 
$N\geq4$, i.e. by requiring at least four identified particles recorded by CHIMERA. 
An important factor in the $Q$-vector definition is the weight factor $\omega = +1(-1)$ for emissions 
in the forward (backward) hemisphere in the c.m. system. It is omitted here because emissions in the 
forward hemisphere are exclusively selected with the condition on rapidity $y_{c.m.}>0.1$.
The vector $\vec{Q}_{\rm CHI}$ represents a $Z$- and transverse-velocity-weighted, 
i.e. approximately transverse-momentum-weighted, direction in the plane perpendicular to the beam direction.

In the AToF analysis, a second vector $\vec{Q}_{\rm AToF}$ has been determined from the recorded positions of 
the interaction of detected fragments with the Time-of-Flight Wall. The horizontal coordinates were 
determined with the uncertainty given by the slat widths of 2.5 cm. It reduces to 1.25 cm if the 
fragment was identified in both layers as observed in most cases. The vertical coordinate was 
determined from the measured difference of the top and bottom time signals, and a resolution of 
typically about $\pm 2$~cm was obtained. The distance to the beam axis, under the assumption of 
approximately beam velocity, is proportional to the transverse velocity of the detected particle or fragment. 
The resulting azimuthal vector was weighted with the atomic 
number $Z$ of the fragment and $\vec{Q}_{\rm AToF}$ was obtained by summing over all individual vectors 
within an event. Also here, the weight factor $\omega$ can be omitted as the AToF acceptance of 
$\theta_{\rm lab} \le 7^{\circ}$ strongly favors projectile fragments. A time-of-flight gate
selecting forward emissions in the c.m. frame was used in addition. 

\begin{figure}[!t]            
\includegraphics[width=7.7cm]{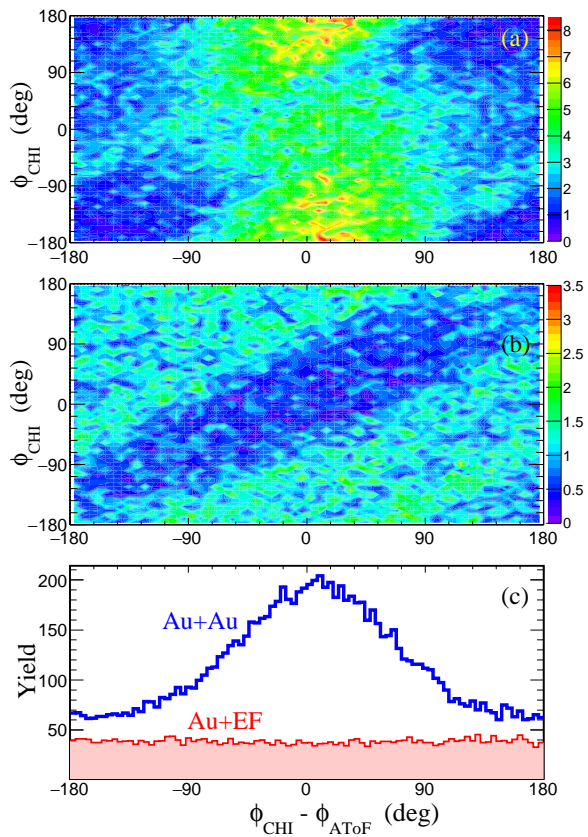}
\caption{Bidimensional representations of the difference of the azimuthal reaction-plane orientations 
individually obtained from CHIMERA and the AToF Wall, under the condition that the CHIMERA and Microball orientations are within 
the adopted anticorrelation gate and shown for measurements with (a) and without (b) a gold foil in the target frame.
Panel (c) shows yield curves for these two cases, Au+Au and Au+EF (EF stands for empty frame), normalized with
respect to the integrated beam intensity.
}
\label{fig:azitofchim} 
\end{figure}

The resolution obtained with these two quantities is overall comparable but depends somewhat on the 
impact parameter. Peripheral collisions associated with small multiplicities in the CHIMERA part of 
the recorded event may be more easily characterized with the heavy fragments seen in AToF while more central 
collisions leading to high CHIMERA multiplicities may produce only few light fragments within the 
acceptance of the AToF Wall. As it turned out, in the impact parameter range of interest, 
central with $b \le 7.5$~fm, only about 10\% of the events permitted the calculation of a $Q$ vector from AToF
hits alone. Because the AToF geometry is not azimuthally symmetric, the resulting 
inclusive $Q$-vector distributions are not fully isotropic.

With the Microball data, the reaction-plane orientation was estimated by summing over the azimuthal 
directions of the recorded hits. 
A vector $\vec{Q}_{\rm \mu Ball}$ has been calculated as
\begin{equation}
\vec{Q}_{\rm \mu Ball}= \sum_{i=1}^{N} \hat{r}_{t}^{i}, 
\end{equation}
where $\hat{r}_{t}^{i}$ is the azimuthal unit vector in the direction of the location of the 
detector module that recorded the $i^{th}$ hit. A minimum multiplicity of $N\geq2$ was imposed. 
In this case, the weight factor $\omega$ has been omitted because the rapidity of the detected 
particles was not determined even though the Microball acceptance of $\theta_{\rm lab} \ge 60^{\circ}$ 
can be expected to select mainly backward emissions. As shown below, the orientation of
$\vec{Q}_{\rm \mu Ball}$ was indeed found to be opposite to those of the CHIMERA and AToF 
$Q$ vectors.

\begin{figure}[!t]            
\includegraphics[width=8.5cm]{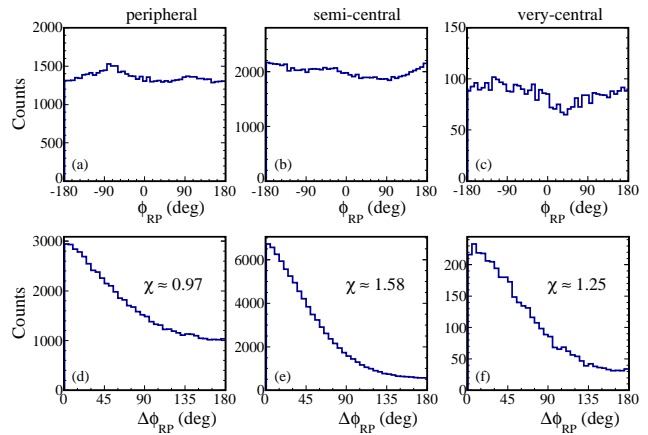}
\caption{(Top row) Inclusive distributions of the angle $\Phi_{\rm RP}$ representing the reaction-plane orientation 
obtained with the $Q$-vector method from the combined CHIMERA and AToF data for peripheral (a), 
semicentral (b), and very central (c) impact-parameter intervals (see Table~\ref{tab:impact} for their definitions). \\
(Bottom row) Distributions of the difference of orientations of the subevent reaction planes for the same three event classes, 
peripheral (d), semicentral (e), and very central (f), obtained with the mixing 
technique of Refs.~\protect\cite{Ollxx,Oll98} and by using the weight $Z\beta_{t}\gamma$ (see text and Table~\ref{tab:orient}). 
The corresponding values of the reaction-plane dispersion 
parameter $\chi$ are indicated. 
}
\label{fig:azireso} 
\end{figure}

\begin{figure}[t]            
\centering
\includegraphics[width=8.5cm]{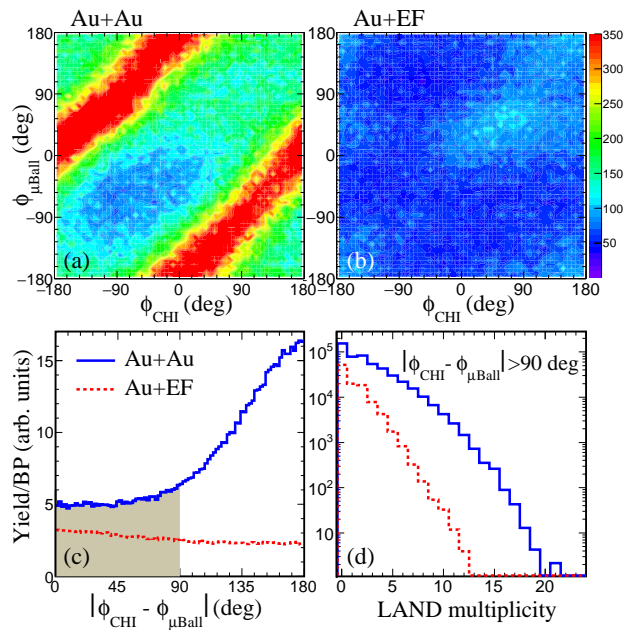}
\caption{(Top row) Correlation between the $Q$-vector orientations determined with CHIMERA 
(abscissa) and with the Microball (ordinate) for data sets taken with (a)) and without (b) a 
target foil in place (EF stands for empty frame). \\
(Bottom row) Difference of the $Q$-vector orientations for Au+Au 
and for Au+EF data (c), normalized with respect to the integrated beam intensity (BP 
stands for beam particles), and the raw hit multiplicities (d) registered with LAND 
for Au+Au (solid line) and for Au+EF data sets (dotted line). The hatched area in (c) indicates the
range of events rejected by the required anticorrelation of the CHIMERA and Microball $Q$-vector   
orientations (Sec.~\protect\ref{sec:corr}).
}
\label{fig:ef}       
\end{figure}

The three $Q$-vectors are strongly correlated. The degree of coincidence of the azimuthal orientations of the 
vectors $\vec{Q}_{\rm CHI}$ and 
$\vec{Q}_{\rm AToF}$ for the class of events containing a valid $\vec{Q}_{\rm AToF}$ is shown in Fig.~\ref{fig:azitofchim}. The individual reaction-plane 
orientations obtained from the CHIMERA and AToF Wall data are evidently very similar. With the target foil removed [panel (b)], the coincidence of orientations is no longer present; the 
correlation pattern is dominated by the slightly reduced acceptance of AToF in the region near 0$^{\circ}$.
The resulting distributions of the difference 
$\Phi_{\rm CHI} - \Phi_{\rm AToF}$ is shown in the
bottom panel. The azimuthal angle $\Phi$ that is used here and in Figs.~\ref{fig:azireso} and~\ref{fig:ef}
is defined in accordance with the chosen coordinate system (Fig.~\ref{fig:setup}), with
$\Phi = 0^{\circ}$ coinciding with the $x$ and $\Phi = 90^{\circ}$ with the $y$ 
direction. 
The applied condition requiring that the CHIMERA and Microball orientations are within the adopted 
anticorrelation gate of $\pm 90^{\circ}$ suppresses unwanted background, as discussed in Sec.~\ref{sec:corr} in more detail. 

The inclusive reaction-plane distributions, 
as given by the combined $Q$-vectors obtained by summing over recorded hits in CHIMERA and AToF 
for three choices of impact-parameter windows, are shown in the top row of 
Fig.~\ref{fig:azireso}. The observed flatness indicates that the particle angular distributions have not been 
biased by variations of the detector 
efficiencies, by properties of the event triggering or by other azimuthal asymmetries in the experiment. 

Several different methods 
of estimating the reaction-plane orientation were applied to the data to identify possible
systematic uncertainties related to it. They are all based on the $Q$-vector method of Ref. \cite{Dan85}
but differ in the kinematic quantities used as weights for summing over the included particles and fragments. 
Besides the product $Z\beta_{t}\gamma$ [cf. Eq.~(\ref{eq:qvec})], equal weights for all particles and 
the atomic number $Z$ alone of each particle were also used as weights for 
summing over the azimuthal directions of the recorded hits either in both CHIMERA and AToF
or in CHIMERA alone. It was, in addition, 
investigated as to what extent the result varies with the value of the rapidity gate 
chosen for selecting the forward hemisphere in the c.m. reference frame. 

\begin{table}[h]
\begin{tabular}{|c|c|c|c|c|c|}
\hline
Detectors and chosen weight              & $y_{c.m.} > 0.1$ & $y_{c.m.} > 0.2$ \\ \hline
CHIMERA alone, equal weight                   &       1.39       &    1.30          \\ \hline
CHIMERA+AToF, equal weight                    &       1.45       &    1.37          \\ \hline
CHIMERA alone, $Z$                            &       1.51       &    1.42          \\ \hline
CHIMERA+AToF, $Z$                             &       1.58       &    1.50          \\ \hline
CHIMERA alone, $Z\beta_{t}\gamma$             &       1.52       &    1.42          \\ \hline
CHIMERA+AToF, $Z\beta_{t}\gamma$              &       1.59       &    1.49          \\ \hline
\end{tabular}
\caption{Resolution parameter $\chi$ obtained for the estimation of the reaction-plane orientation with different 
choices for the $Q$-vector construction for the case of semicentral $^{197}$Au+$^{197}$Au collisions. 
The first column indicates the considered detector 
systems and weights; the second and third columns show $\chi$ for two values of the rapidity gate chosen for CHIMERA hits. 
}
\label{tab:orient}
\end{table}

The criterion chosen for this investigation was the achieved resolution of the reaction-plane orientation. 
It determines the necessary corrections and the uncertainty associated with the obtained flow parameters~\cite{And06}.
It was evaluated with the subevent mixing 
technique as described in Refs.~\cite{Ollxx,Oll98} and quantified through the resolution parameter $\chi$.
This parameter is inversely proportional to the width of the difference distribution of subevent orientations,
assumed to be Gaussian in the present case (cf. Ref.~\cite{And06}).
Examples of difference distributions obtained for selected intervals of impact parameter 
are given in the bottom row of Fig.~\ref{fig:azireso}, including the corresponding results for $\chi$. 
The resolution parameters obtained with the studied choices of weights and detector systems 
are listed in Table~\ref{tab:orient} for the class of semicentral events. 
The best resolution, indicated by the largest value for $\chi$, has been achieved using the product 
$Z\beta_{t}\gamma$ 
as the weight and by summing over the recorded hits with $y_{c.m.}>0.1$ in both CHIMERA and AToF. 
All the results shown in the following sections were obtained with this choice. 
It is interesting to note, however, that other choices
for the weighting factors lead to very comparable results (Table~\ref{tab:orient}). 

The correction factors resulting from the so determined dispersion of the reconstructed
reaction plane were obtained according to Ref.~\cite{Ollxx,Oll98}. Resolution parameters $\chi$ in the 
range of 1.2 to 1.6 (Fig.~\ref{fig:azireso}) correspond to attenuation factors 
$\langle {\rm cos} (n \Delta\phi) \rangle$ of approximately 0.8 to 0.9 for $n=1$, i.e., for the case of 
directed flow, and from $\approx$ 0.5 to 0.65 for the elliptic flow ($n=2$). Their inverse values
represent the correction factors to be applied to the Fourier coefficients describing the 
measured azimuthal anisotropies. The validity of the method used for determining the reaction-plane
orientation and its experimental dispersion were confirmed by a comparison of collective flows obtained
from the KRATTA and from FOPI data~\cite{reisdorf2013} for the same reaction. Excellent agreement is obtained 
for directed and elliptic flows of hydrogen and helium isotopes within the common acceptance of the two 
experiments~\cite{kupny2014}.

\subsection{Background corrections}
\label{sec:corr}

For rejecting background reactions owing to the interaction of Au projectiles with non-target material, 
the correlation of the $Q$-vector orientations as given by CHIMERA and by the  
Microball detectors was used.
Figure~\ref{fig:ef} shows the correlation between their azimuthal directions, $\Phi_{\rm CHI}$ and $\Phi_{\rm \mu Ball}$, 
for $^{197}$Au+$^{197}$Au reactions (a) and $^{197}$Au+empty frame (b) data, normalized 
relative to each other with respect to the integrated beam intensities. The strong anticorrelation 
for on-target reactions is evident. It is expected because forward-emitted particles were selected with CHIMERA ($y_{\rm c.m.} > 0.1$)
and the Microball covers mainly the backward hemisphere in the c.m. frame. 

In runs with empty target frames, the recorded 
yields are low and only a weak positive correlation is observed. The distribution of differences between the two $Q$-vector 
orientations, normalized with respect to the integrated beam intensity, is presented in panel (c). 
To minimize the contributions of non-target collisions in the data analysis, an anticorrelation 
of the CHIMERA and Microball $Q$-vector orientations was required. The applied 
condition $\vert\Phi_{\rm CHI}-\Phi_{\rm \mu Ball}\vert>90^{\circ}$ led to a relative weight of background reactions of less than 20\%. 
It underlines the importance of the Microball data for identifying and rejecting off-target reactions. 

Panel (d) of Fig.~\ref{fig:ef} shows the LAND raw multiplicity (number of modules hit per event), 
normalized with respect to the integrated beam intensity, for $^{197}$Au+$^{197}$Au and $^{197}$Au+empty 
frame data and after applying the CHIMERA-Microball anticorrelation condition. The contribution from 
non-target backgrounds in the kinematic region of LAND is weak, starting with less than 20\% at 
unit multiplicity to much less than 1\% at multiplicity 10. In the final analysis,
normalized yields of the remaining non-target background events were subtracted from the corresponding
$^{197}$Au+$^{197}$Au data sets.

\section{Experimental results}
\label{sec:resu}

Azimuthal distributions of neutrons and light-charged particles measured with LAND with respect 
to the reaction plane determined with the CHIMERA and AToF detectors, as described in the previous section, 
were extracted for $^{197}$Au+$^{197}$Au reactions from data collected with and without a target and
without and with the shadow bar in front of LAND. 
After subtracting the measured and normalized background yields, the obtained distributions were 
fitted with the Fourier expansion 
\begin{equation}
f(\Delta\phi) \propto 1 + 2 v_1 {\rm cos}(\Delta\phi) +
2 v_2 {\rm cos}(2 \Delta\phi)
\label{eq:fourier}
\end{equation}
to determine the coefficients describing the observed directed ($v_1$) and elliptic ($v_2$) flows.
$\Delta\phi$ represents the azimuthal angle of the momentum vector of an 
emitted particle with respect to the determined reaction plane~\cite{And06}. 
Owing to insufficient resolution, charge identification with 
the $\Delta$E-vs-time-of-flight technique has not been possible with LAND in the present experiment. 
Therefore, results only for 
neutrons and for all recorded charged particles are presented in the following.

\begin{figure}[!t]           
\centering
\includegraphics[width=8.0cm]{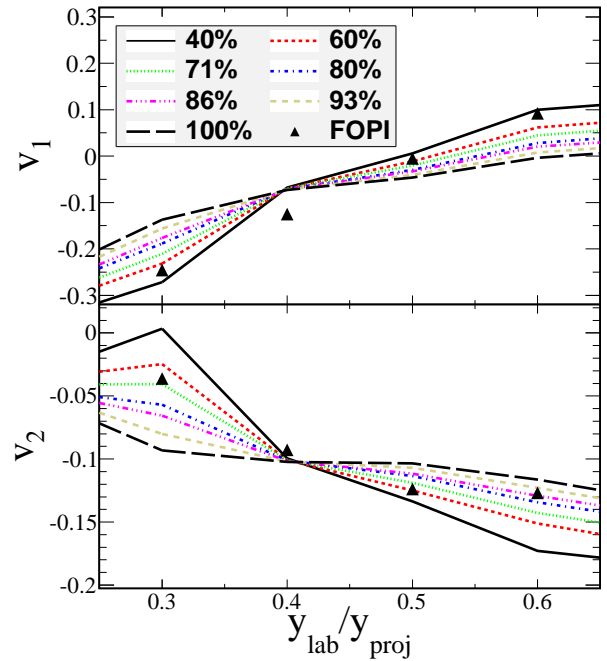}
\caption{Measured directed (top) and elliptic flows (bottom) of charged particles as determined with different 
timing corrections in comparison with FOPI results (solid triangles, from~\protect\cite{reisdorf2013}) for the same $^{197}$Au+$^{197}$Au reaction at 400 MeV/nucleon in the interval of impact parameters $3.35 \le b \le 6$~fm. The percentages of cases to which the so-called second step of 
the timing corrections was applied are given in the legend (see text). 
The solid and dashed black lines indicate the limits 40\% and 100\%, respectively, of the 
studied probability interval.
}
\label{fig:fopiflow}       
\end{figure}

\subsection{Timing corrections}
\label{sec:timcorr}

The timing information of particles detected with LAND in these data sets had been corrected as described in
Sec.~\ref{sec:landtiming} and in the Appendix. 
One of the unknown parameters appearing in this procedure was the number 
of particles misplaced or 
wrongly corrected in the so-called second step. Therefore, a series of analysis runs was performed in which 
the percentage of particles subjected to it was reduced from 100\% to 40\% in steps of increasing width. 
The resulting flow parameters are shown in Fig.~\ref{fig:fopiflow} as a function of the reduced 
rapidity $y_{\rm lab}/y_{\rm proj}$. It is observed that the influence of the second correction
is negligible at rapidities $y_{\rm lab}/y_{\rm proj} \approx 0.4$ but significant at lower and higher 
rapidities. At a reduced rapidity $y_{\rm lab}/y_{\rm proj} = 0.4$, 
the acceptance of LAND in this experiment selects transverse momenta 
of approximately 0.3 to 0.5 GeV/$c$/nucleon for which the discussed effect is, apparently, less severe.
As expected for a mixing between time intervals, the modifications at low and high rapidities 
occur in opposite directions for both the directed and the elliptic flows. 

For the data selected for this purpose, an interval of nominal impact parameters
$3.35 \le b \le 6$~fm was chosen because corresponding flow data have been made available by the
FOPI Collaboration~\cite{reisdorf2013}. 
It is contained within the semicentral event class and its parameters are listed in the bottom line of Table~\ref{tab:impact}.
The comparison is not meant to identify a ``best'' percentage at which 
the problem will largely disappear. It only shows that the 100\% application of the second step does not necessarily
lead to improved flow values, consistent with the observation made for the time spectra discussed in the Appendix.
It also suggests an application with 40\% as a useful lower limit.
Variations within this interval of 40\% to 100\% are considered as suitable for quantifying the contribution of the
mixing and the underlying timing uncertainty to the systematic error of the measurement. 
It applies mainly to the flow parameters deduced as a function of rapidity or of transverse momentum. 
The effect is of minor importance for the acceptance-integrated flow ratios 
based on time-integrated particle yields.

\subsection{Collective flow}   
\label{sec:collflows}

\begin{figure}[!t]           
\centering

\includegraphics[width=8.0cm]{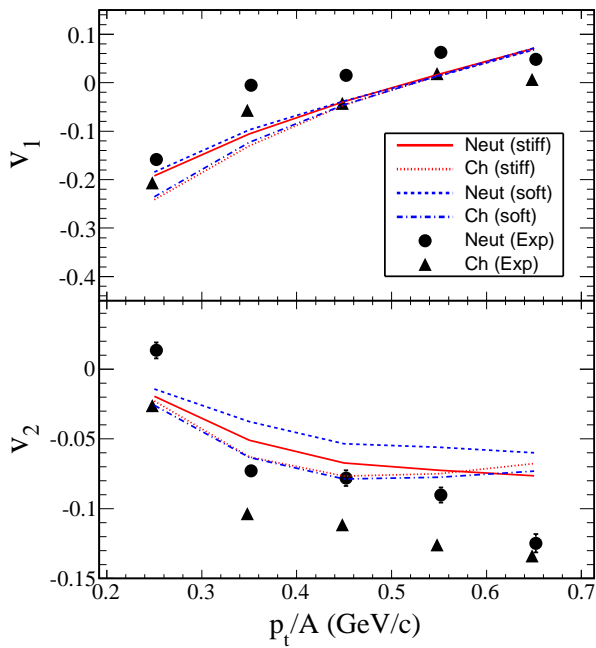}
\caption{Measured flow parameters $v_1$ (top) and $v_2$ (bottom) for the central event class ($b<7.5$~fm) in $^{197}$Au+$^{197}$Au 
collisions at 400 MeV/nucleon for 
neutrons (solid circles) and charged particles (solid triangles) as a function of the transverse
momentum $p_t/A$.
The UrQMD predictions for neutrons and charged particles obtained with a stiff ($\gamma$=1.5, red solid and dotted 
lines, respectively) and a soft ($\gamma$=0.5, blue dashed and dash-dotted lines, respectively) density dependence 
of the symmetry term have been filtered to correspond to the geometrical acceptance of the experiment. 
The experimental data are corrected for the dispersion of the reaction-plane orientation. 
Where not shown, the statistical errors are smaller than the size of the symbols.
}
\label{fig:flows}       
\end{figure}

Flow parameters obtained after correcting for the dispersion of the reaction plane are shown in Fig.~\ref{fig:flows} 
as a function of the transverse momentum per particle $p_t/A$.
They are integrated over the rapidity range covered by the LAND acceptance which increases with $p_t/A$
from $y_{\rm lab}/y_{\rm proj} \approx 0.3$ to 0.7 (cf. Fig. 1 of Ref.~\cite{Rus11}). 
The observed yield of particles decreases rapidly with increasing 
transverse momentum, so that the low-intensity regions at high $p_t$ are more strongly
affected by occasionally misplaced particles originating from the regions of high yield at lower $p_t$. 
For this reason, the analysis is restricted to transverse momenta $p_t/A \le 0.7$~GeV/$c$.
The selected range of nominal impact parameter is $b \le 7.5$~fm (central event class), and a fraction of 80\% is chosen for the 
application of the second correction step discussed above, compatible with the comparison of elliptic-flow results
shown in Fig.~\ref{fig:fopiflow}. 
The coefficient $v_1$ rises from negative
values for small $p_t/A$ to small positive values at $p_t/A > 0.6$, reflecting the correlation
of transverse momentum with rapidity caused by the acceptance of LAND. 
The coefficient
$v_2$ is small at small $p_t/A$ and assumes values below $v_2 = -0.1$ at larger $p_t/A$, indicating the 
strength of particle squeeze-out in the directions perpendicular to the reaction plane.

\section{Interpretation with UrQMD}
\label{sec:urqmd}

As in the earlier FOPI-LAND study~\cite{Rus11}, the ultrarelativistic QMD (UrQMD) model  
of the group of Li and Bleicher~\cite{qli05,qli06,Li:2006ez} has been employed to deduce
the density dependence of the nuclear symmetry energy. Even though alternative parametrizations have recently become 
available~\cite{wang14a,wang14b}, the version employed in the FOPI-LAND study was used again, so as to
permit a direct comparison of the density dependencies obtained from the two experiments.
The differences are, furthermore, not very large.
In the study presented by Wang {\it et al.} using a variety of Skyrme forces a very comparable stiffness
parameter $L = 89 \pm 23$~MeV was obtained, differing from the original result $L = 83 \pm 26$~MeV 
by only a few MeV~\cite{Rus11,wang14b}. 
The parameter 
\begin{equation}
L = 3\rho_0 \frac{\partial E_{\rm sym}}{\partial \rho}|_{\rho=\rho_0} 
\label{eq:L}
\end{equation}
is proportional to the slope of the symmetry energy at saturation (see, e.g., Ref.~\cite{lipr08}).

The UrQMD model was originally developed to study particle production at high 
energy~\cite{bass98}. By introducing a nuclear mean field 
with momentum-dependent forces, it has been adapted to the study of intermediate-energy heavy-ion collisions~\cite{qli09}.
The chosen equation of state is soft. The updated Pauli-blocking scheme, introduced to provide 
a more precise description of experimental observables at lower energies, is described in Ref.~\cite{qli11}.
Different options for the dependence on isospin asymmetry were implemented. Two of them are used here, 
expressed as a power-law dependence of the potential part of the symmetry energy on the
nuclear density $\rho$ according to
\begin{equation}
E_{\rm sym} = E_{\rm sym}^{\rm pot} + E_{\rm sym}^{\rm kin} 
= 22~{\rm MeV} (\rho /\rho_0)^{\gamma} + 12~{\rm MeV} (\rho /\rho_0)^{2/3},
\label{eq:pot_term}
\end{equation}
with $\gamma =0.5$ and $\gamma =1.5$ corresponding to a soft and a stiff density 
dependence.  

The UrQMD predictions for these two choices are shown in Fig.~\ref{fig:flows} in comparison with the experimental data 
for both neutrons and charged particles. 
A filtering procedure was used to adapt the results to the experimental conditions. 
They qualitatively follow the experimental flow values, even though the predicted 
squeeze-out is less pronounced than that observed. A significant
sensitivity with respect to the stiffness of the symmetry energy is visible for the elliptic flow of 
neutrons. By comparing it to the strength of the charged-particle flow in the form of flow ratios or 
differences, this sensitivity is expected to be preserved,
even in the presence of a global over- or underprediction of the elliptic flows~\cite{Rus11,Coz11}.

The slight underprediction is known to be related to 
the so-called FP1 parametrization for the momentum dependence of the elastic nucleon-nucleon cross sections 
in the default version of the UrQMD model that was used here. 
UrQMD studies of the reaction dynamics at intermediate energies have shown that 
the in-medium modification of the elastic nucleon-nucleon cross-sections is an important 
ingredient for realistic descriptions, and various parametrizations have been tested~\cite{wang14a}. 
In the previous FOPI-LAND study, additional calculations were performed with the FP2 parametrization, 
causing the elliptic-flow parameter $v_2$ to be slightly overpredicted.
The absolute values of $v_2$ obtained with FP1 and FP2 differ by $\approx 40\%$ for this reaction~\cite{qli11,qli10}. 
The calculated ratios retain, nevertheless, the sensitivity of the elliptic flow 
to the stiffness of the symmetry energy and depend only weakly on the chosen
parametrization for the in-medium cross sections~\cite{Rus11}.

The systematic study of the residual model dependence of transport descriptions of the elliptic-flow ratios 
and differences by Cozma {\it et al.}~\cite{Coz13} has, in addition, demonstrated that the T\"{u}bingen QMD transport 
model used there leads to equivalent results regarding the deduced stiffness of the symmetry energy. In particular, also
the impact of including or neglecting the momentum dependence of the symmetry potential was investigated
with different parametrizations. Important input quantities identified by this study were the isoscalar compressibility 
and the width of the nucleon wave function employed in the calculations.
Narrower constraints for these quantities will reduce the theoretical uncertainties.
A quantitative study of the model differences between the UrQMD and the T\"{u}bingen versions was performed by 
Wang {\it et al.}~\cite{wang14b}. Expressed in terms of the central value obtained for the slope parameter $L$, an uncertainty 
of $\Delta L \approx 10$~MeV may be ascribed to the observed model dependence of the UrQMD
versus the T\"{u}bingen-QMD analyses. 

Besides the momentum-dependence of the symmetry potential~\cite{Das:2002fr,baoan_mdi04,Gio10,feng12,leizhang12,Xie15},
attention has to be paid to the recent observation of short-range correlations~\cite{subedi08,hen14sc}, 
leading to larger tails of the momentum distributions in symmetric matter than in pure neutron matter and
to a reduction of the kinetic part in the parametrization 
of the symmetry energy~\cite{carbone12,rios14,liguoshi15}. It will be interesting
to incorporate these correlations in transport models and to explore their consequences~\cite{hen15,caili16}. 
However, in a first study~\cite{yong16}, the effect for elliptic-flow ratios was found to be negligibly small
for the case of a mildly soft to linear density dependence of the symmetry energy that is supported by the present data.
It is, nevertheless, evident that the improvement of current theoretical descriptions is an important goal for the future.
Reducing theoretical uncertainties and enhancing their consistency~\cite{junxu16}, will permit tighter constraints 
for the high-density dependence of the symmetry energy.

The UrQMD transport program is stopped at a collision time of 150 fm/$c$ and a conventional phase-space coalescence
model with two parameters is used to construct clusters. Nucleons 
with relative momenta smaller than $P_0$ and relative distances smaller
than $R_0$ are considered as belonging to the same cluster. The values $P_0 = 0.275$~GeV/$c$
and $R_0 = 3.0$~fm have been adopted as standard parameters.  
With these values the overall dependence of cluster yields on $Z$ is rather well
reproduced but the yields of $Z=2$ particles are underpredicted~\cite{Rus11}. 
In the comparison with the FOPI data set used for Fig.~\ref{fig:fopiflow}, after normalization with respect to $Z=1$,
an underprediction by a factor 1.4 was observed. 
The yields of deuterons and tritons in central 
collisions are also underestimated by similar factors. 

Constraints for the symmetry energy were determined by comparing the ratios 
of the elliptic flows of neutrons and charged particles (ch), $v_2^{n}/v_2^{ch}$, with the corresponding UrQMD 
predictions for the soft and stiff assumptions. Because hydrogen isotopes could not be selected, as done in the
FOPI-LAND study~\cite{Rus11}, a test was performed for confirming the equivalence of results obtained
when including all recorded charged particles in the analysis. For this purpose, the data of the FOPI-LAND 
experiment were analyzed with and without the condition $Z=1$ applied in the charged-particle selection
and with the limitation $p_{t}/A \le 0.7$~GeV/$c$ of the integration interval in transverse momentum. 
The corresponding power-law coefficients $\gamma$ were determined by comparing with UrQMD calculations 
performed with the same conditions. 
In addition, the effect of enhancing the weight of the
$Z=2$ contribution to the calculated $Z$-integrated flow was tested. Because good agreement was obtained with an enhancement factor 1.4, corresponding to the observed underprediction, it was used as default option in the analysis. Overall, the changes observed in these tests for the central values were less than
$\Delta \gamma =0.05$, accompanied however by the larger statistical error of the FOPI-LAND data set.

\begin{figure}[t]           
\centering
\includegraphics[width=8.0cm]{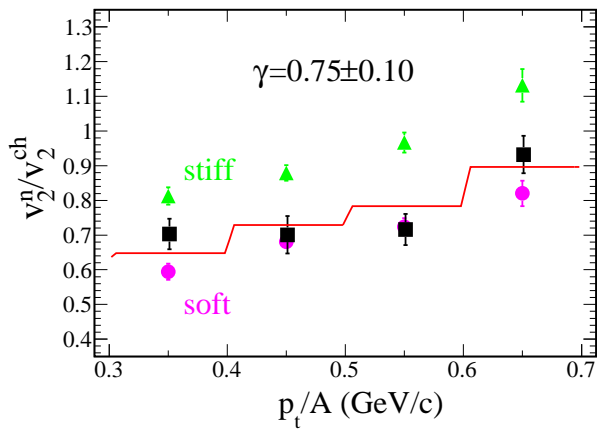}
\caption{Elliptic flow ratio of neutrons over all charged particles for central ($b<$ 7.5 fm) 
collisions of $^{197}$Au+$^{197}$Au at 400 MeV/nucleon
as a function of the transverse momentum per nucleon $p_{t}/A$, evaluated with a fraction of 80\% for the
second step of timing corrections (see Sec.~\protect\ref{sec:timcorr}). 
The black squares represent the experimental data; the green triangles and purple circles represent the UrQMD predictions
for stiff ($\gamma =1.5$) and soft ($\gamma =0.5$) power-law exponents of the potential term, respectively. 
The solid line is the result of a linear interpolation between the predictions, weighted according to the experimental errors of the included four bins in $p_{t}/A$ and leading to the 
indicated $\gamma =0.75 \pm 0.10$.
}
\label{fig:diffdata}       
\end{figure}

\subsection{Differential data}
\label{sec:diffdat}

The ratio $v_2^{n}/v_2^{ch}$ obtained from the present data for the class of central ($b<$ 7.5 fm) 
collisions as a function of the transverse momentum per nucleon $p_{t}/A$ is shown in 
Fig.~\ref{fig:diffdata}. The chosen fraction for the second step of timing corrections 
(see Sec.~\protect\ref{sec:timcorr}) is 80\%, compatible with the comparison with FOPI data presented 
in Fig.~\ref{fig:fopiflow}.  
Under this assumption, the best description of the neutron-vs-charged-particle elliptic flow is
obtained with a power-law coefficient $\gamma=0.75\pm 0.10$, where $\Delta\gamma=0.10$ is the statistical
uncertainty returned by the fit routine. It results from linearly interpolating between 
the predictions for the soft, $\gamma=0.5$, and the stiff, $\gamma=1.5$, predictions of the model
within the range of transverse momentum $0.3 \le p_{t}/A \le 0.7$~GeV/$c$.

\begin{figure}[t]           
\centering
\includegraphics[width=7.5cm]{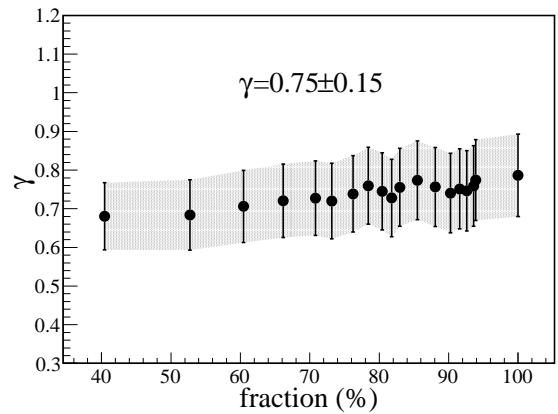}
\caption{
Potential-term coefficient $\gamma$ deduced by interpolating between the UrQMD predictions shown in 
Fig.~\protect\ref{fig:diffdata} as a function of the fraction chosen for the second step of 
timing corrections (see Sec.~\protect\ref{sec:timcorr}).  
}
\label{fig:fraction}       
\end{figure}

The dependence of the resulting $\gamma$ on the choice made for the second timing correction in the data analysis is
shown in Fig.~\ref{fig:fraction}. 
Under the assumption that the second correction should be applied to at least 40\% of the corresponding particles, 
the 1-$\sigma$ error margins are confined within the interval $\gamma=0.75\pm 0.15$ as apparent from the figure. 
The larger error $\Delta\gamma=0.15$ is expected to include the systematic uncertainty caused by the existence
of misplaced hits, not identified in the first step and only partly included in the second step of the timing 
correction scheme of the analysis.

\subsection{Acceptance-integrated flow ratio}
\label{sec:integ}

\begin{figure}[t]           
\centering
\includegraphics[width=8.5cm]{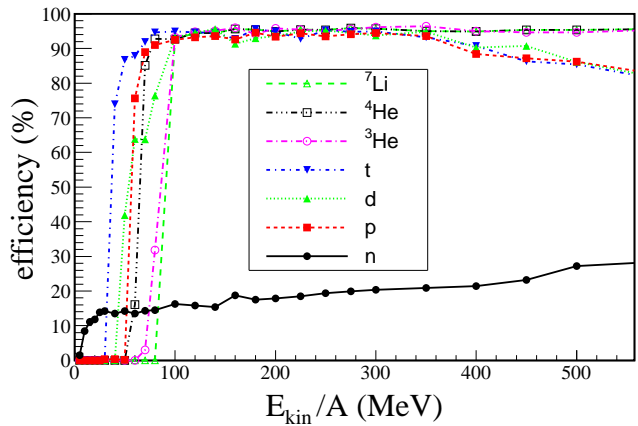}
\caption{Detection probability of the first plane of the LAND detector, preceded by the veto wall, for neutrons (dots), protons
(solid squares), deuterons (solid triangles), tritons (solid tip-down triangles), $^3$He (open circles),
$^4$He (open squares), and $^7$Li (open triangles) as a function of the particle kinetic energy per 
nucleon $E_{\rm kin}/A$.
}
\label{fig:eff}       
\end{figure}

The new constraint deduced in the preceding section is slightly lower but still within the uncertainty interval
of the previous value $\gamma=0.9\pm 0.4$ deduced from the FOPI-LAND data and the same UrQMD 
model~\cite{Rus11}. The error is significantly reduced by a factor of more than two. 
To confirm the validity of the obtained result and to minimize complications arising from 
the time-of-flight measurement with LAND, an acceptance-integrated flow ratio was  
determined by integrating over the full $t_{\rm hit}$ spectrum shown in Fig.~\ref{fig:corrb} in the Appendix. 
It includes all recorded particles irrespective of their actual location within this spectrum. The corresponding 
UrQMD calculations were integrated over the full acceptance of LAND as given by the
covered interval of laboratory angles. The thresholds and the energy and particle-type dependent detection 
efficiency of the effectively used first plane of the LAND detector behind the veto wall were taken into account
(Fig.~\ref{fig:eff}). The efficiency calculations were carried out with Geant3 within the FairRoot software 
framework~\cite{fairroot}.

The still remaining uncertainty arising from this procedure is connected with the choice of the upper 
limit of the time-of-flight interval which determines the lower threshold of the neutron energy.
For protons to pass through the veto wall and to be detected in a LAND module, a minimum energy of about
60 MeV is required while neutrons with lower energies may still be detected (Fig.~\ref{fig:eff}). 
The magnitude of this effect has been assessed by varying the upper limit of time-of-flight integration
between 60 and 90~ns, resulting in a slight variation of the obtained flow ratio and the exponent $\gamma$.
The UrQMD calculations were performed for this purpose with kinetic-energy thresholds that corresponded
to the chosen integration limit for neutrons and the physical lower detection thresholds 
for charged particles.
Acceptance-integrated elliptic-flow values were then determined from the azimuthal anisotropy of the 
obtained yields and the linear interpolation between the predictions was used to determine the corresponding
exponents $\gamma$.

\begin{figure}[t]           
\centering
\includegraphics[width=7.5cm]{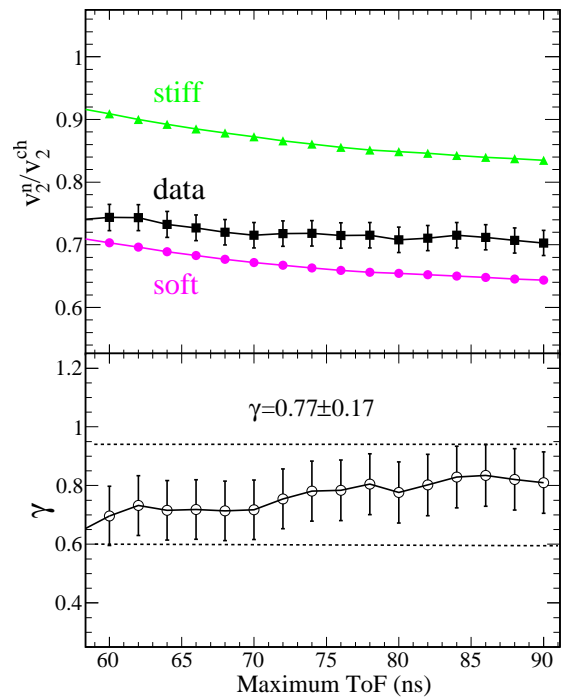}
\caption{Measured elliptic-flow ratio for central ($b<$ 7.5 fm) 
collisions of $^{197}$Au+$^{197}$Au at 400 MeV/nucleon in comparison with stiff and soft UrQMD predictions (top) 
and deduced symmetry term coefficient $\gamma$ (bottom) as a function of the upper limit of the time-of-flight 
interval used to obtain time integrated results. The dashed horizontal lines mark the upper and lower limits of
the 1-$\sigma$ statistical error margin $\Delta\gamma = \pm 0.10$ within the time interval $60 < ToF < 90$~ns. 
}
\label{fig:tofinteg}       
\end{figure}

The results for the measured and calculated acceptance-integrated flow
ratios and the resulting $\gamma$ are shown in Fig.~\ref{fig:tofinteg}. A small monotonic variation of
$\gamma$ with the assumption regarding the upper limit of the ToF interval is evident. The
1-$\sigma$ error margins are confined to the interval $\gamma=0.77\pm 0.17$. 
It overlaps with the interval obtained by varying the fraction of hits included in the second correction 
step (Fig.~\ref{fig:fraction}). 
This is not unexpected as the two methods
are both aiming at quantifying the remaining consequences of not recognized simultaneous timing errors 
of the two signals from a paddle. The variation of the maximum of the $ToF$ interval, 
in addition, includes the effect of a possible smearing of the energy threshold for neutrons and charged 
particles by the 25-ns time jumps.

\subsection{Final corrections}
\label{sec:fincorr}

Up to this point, the effects of charge-changing processes, nuclear or instrumental, have been ignored in the analysis. 
The largest effects of this kind are caused by misidentifications of charged particles as neutrons, because of 
a missing veto signal, and of neutrons as charged 
particles  
because of a neutron-induced reaction in a veto panel 
that produces a signal.
Nuclear charge-exchange reactions with cross sections on the level of millibarn are less important in comparison 
(see, e.g., Refs.~\cite{anderson96,tanihata15}). Furthermore,
protons converted into neutrons in the veto wall may still have left a signal there while neutrons
converted to protons are included in the measured, rather small, efficiency for neutron detection of
the thin veto paddles (see below).
Misidentifications reduce the difference between the measured  flow patterns and thus cause a small increase of the 
apparent flow ratio. The resulting symmetry-term coefficient appears stiffer than without these effects.

\begin{figure}[htb]           
\centering
\includegraphics[width=7.5cm]{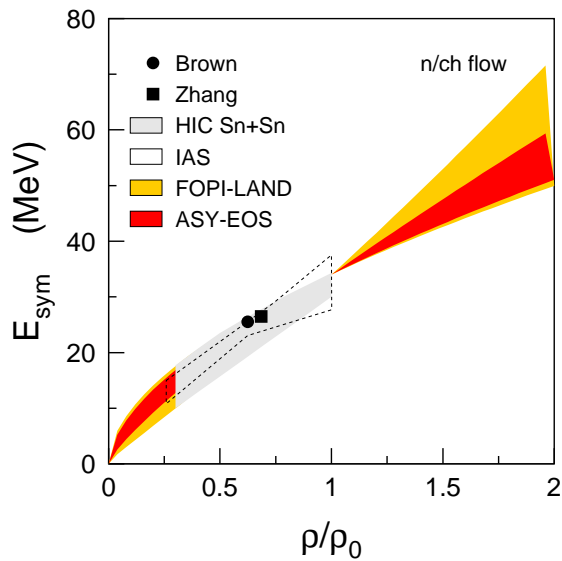}
\caption{Constraints deduced for the density dependence of the symmetry energy from the present data
in comparison with the FOPI-LAND result of Ref.~\protect\cite{Rus11} as a function of the reduced density $\rho/\rho_0$. 
The low-density results of Refs.~\protect\cite{tsang09,dani13,brown13,zhang13} as reported in Ref.~\protect\cite{horowitz13} are 
given by the symbols, the gray area (HIC), and the dashed contour (IAS). For clarity, the FOPI-LAND and ASY-EOS results are
not displayed in the interval $0.3 < \rho/\rho_0 < 1.0$. 
}
\label{fig:final}       
\end{figure}

Calculations within the R3BRoot simulation framework~\cite{fairroot} 
have been performed with different assumptions regarding the detector response and particle recognition.
In particular, the particle-dependent detection thresholds have been taken into account (Fig.~\ref{fig:eff}).
The obtained reduction of the power-law exponent $\gamma$ varied between $\Delta \gamma = -0.03$
and -0.07, with the lower and upper boundaries being obtained with the most extreme assumptions. 

The magnitude of the required correction is, qualitatively, easily understood. 
A 1-mm gap between veto paddles causes an inefficiency
of approximately 1\%. It may cause the equivalent amount of charged particles to appear as 
neutrons in the analysis. 
As charged particles by nature, they have a five-fold higher probability for being detected in the first 
layer of LAND. By taking into account the known
yield ratio of charged particles over neutrons of approximately 2/3 and the measured flow ratio of  
$v_2^n / v_2^{ch} = 0.72$ (Fig.~\ref{fig:tofinteg}), a corrected ratio $v_2^n / v_2^{ch} = 0.71$ is obtained.
With the sensitivity of the flow ratio as represented in the figure, the correction amounts to 
$\Delta \gamma = -0.05$. It represents an upper limit for this particular effect because the veto paddles 
are aligned with respect to the elements of the first plane of LAND and not all particles passing through 
the veto gaps are recorded. As an analysis detail, we note here that in testing the coincidence of timing 
signals in the veto wall and first plane of LAND the possibility of undetected $\pm 25$-ns displacements 
of one of the signals was taken into account (errors in the positions derived from the time signals have all 
been corrected, see Appendix). Other processes exist but are less important. 
The detection probability for neutrons in the 5-mm veto paddles is below
1\% (cf. Fig.~1 of Ref.~\cite{LAND}) and the coincidence requirement of a matching hit in the first module of 
LAND further reduces the probability of misidentifications of this kind. In the simulations, all these effects
are included.

The adopted reduction $\Delta \gamma = -0.05 \pm 0.02$ leads to a final result for the power-law 
coefficient $\gamma = 0.72 \pm 0.19$. The quoted uncertainty is obtained by a linear addition of the  
systematic error of the correction and the $\Delta \gamma = -0.17$ uncertainty resulting from the comparison 
of the acceptance-integrated flow ratio with the UrQMD calculations (Fig.~\ref{fig:tofinteg}). 
The possibility of charge misidentifications considered
here has not been taken into account in the FOPI-LAND analysis. There its magnitude appears small
in comparison with the uncertainty $\Delta \gamma = \pm 0.4$ of this earlier result. It was also not included 
yet in presentations of preliminary ASY-EOS results at conferences~\cite{russotto15}.

The obtained constraint for the density dependence of the symmetry energy is shown in 
Fig.~\ref{fig:final} in comparison with the FOPI-LAND result of Ref.~\cite{Rus11} as a function 
of the reduced density $\rho/\rho_0$. The new result confirms the former and has a considerably smaller 
uncertainty. Judging from the purely statistical error of $\Delta\gamma= \pm 0.10$ (Fig.~\ref{fig:diffdata}),
even smaller errors can be expected from future measurements. 

For reference, the low-density behavior of the symmetry energy from 
Refs.~\cite{tsang09,dani13,brown13,zhang13} as reported in Ref.~\cite{horowitz13} is included in the figure. 
The present parametrization is found compatible also with these results from nuclear structure 
studies and from reactions at lower bombarding energy. 
The corresponding slope parameter describing the variation of the symmetry energy with density at saturation
is $L = 72 \pm 13$~MeV. Judging from the analysis work done with the FOPI-LAND data, one may expect that
the analysis of the present data with the T\"{u}bingen QMD~\cite{Coz11,Coz13} will lead to a similar or 
possibly slightly larger value for the parameter $L$~\cite{Rus14,wang14b,Traut14}. 

The sharp value $E_{\rm sym} (\rho_0) = 34$~MeV is a consequence of the chosen parametrization [Eq.~(\ref{eq:pot_term})]. 
Using values lower than the default $E_{\rm sym}^{\rm pot} (\rho_0) = 22$~MeV, as occasionally done in other UrQMD 
studies~\cite{wang14a,wang2012}, is likely to lower the result for $L$. Values of the symmetry energy at saturation in the
interval between 30 MeV and 32 MeV seem to be favored by a majority of terrestrial experiments and astrophysical 
observations as shown in recent compilations~\cite{lihan2013,lattimer14}. Motivated by these results, the present UrQMD analysis
has, in addition, been performed with $E_{\rm sym}^{\rm pot} (\rho_0) = 19$~MeV, corresponding to $E_{\rm sym} (\rho_0) = 31$~MeV. The obtained power-law coefficient $\gamma = 0.68 \pm 0.19$ is lower by $\Delta\gamma = 0.04$ and the 
corresponding slope parameter $L = 63 \pm 11$~MeV is lower 
by 9~MeV, changes that both remain within the error margins of these quantities.
It is to be noted, however, that the precise results of 
Brown~\cite{brown13} and Zhang and Chen~\cite{zhang13} are no longer met with this alternative parametrization of the symmetry energy.

\section{Density probed}
\label{sec:density}

Calculations predict that central densities of two to three times the saturation density may be reached 
in $^{197}$Au+$^{197}$Au collisions at 400 to 1000 MeV/nucleon on time scales of 
$\approx 10 - 20$~fm/$c$~\cite{li_npa02}. 
The resulting pressure produces a collective outward motion of the compressed material whose strength,
differentiating between neutrons and protons,  
is influenced by the symmetry energy in asymmetric systems~\cite{dani02}. It is to be expected, however,
that the observed transverse momenta of emitted particles and their azimuthal variation
apparent as elliptic flow carry information acquired during the full reaction history. The tests
performed with the FOPI-LAND flow data and varying parameters for the potential part of the symmetry energy
in the UrQMD had already indicated that densities above and below saturation contribute to the observed 
flow patterns~\cite{Rus11}.

A force-weighted density has been defined by Le F\`{e}vre {\it et al.} in their recent study of the 
equation of state
of symmetric matter, based on FOPI elliptic-flow data and IQMD calculations~\cite{lefevre15}. For
$^{197}$Au+$^{197}$Au collisions at 400 MeV/nucleon, its broad maximum extends over densities $0.8 < \rho/\rho_0 < 1.6$.
Liu {\it et al.} report in their study of pion production in the same reaction that the relative sensitivity 
of the $\pi^-/\pi^+$ yield ratios to the symmetry energy is distributed
over a density range of approximately $0.7 < \rho/\rho_0 < 1.8$ with a maximum near 
$\rho/\rho_0 \approx 1.2$~\cite{liuyongwen15}. These more quantitative studies, with partly different 
methods, consistently support the assumption that suprasaturation densities up to nearly twice saturation 
are probed at this energy with
collective flows and meson production, not exclusively but with major effects produced above saturation. 

For the present purpose, a detailed analysis of the collision processes has been performed with 
the T\"{u}bingen version~\cite{Coz13} of the QMD model (T\"{u}QMD). 
The sensitivity to the various density regimes probed in heavy-ion collisions was studied quantitatively
by examining their impact on the variation of elliptical-flow-ratio observables with the two choices of
a mildly stiff and a soft parametrization for the density-dependent asymmetric-matter equation of state (asy-EoS). 
To that end, the density-dependent quantity DEFR (difference of elliptic-flow ratio)
\begin{eqnarray}
{\rm DEFR}^{(n,Y)}(\rho)=\frac{v_2^n}{v_2^{Y}}(x=-1,\rho)-\frac{v_2^n}{v_2^{Y}}(x=1,\rho)
\label{defrdef}
\end{eqnarray}
has been determined using the T\"{u}QMD transport model. Here $Y$ labels a particle or a group of particle species 
and $x$ the asy-EoS stiffness resulting from the momentum-dependent one-body potential introduced by 
Das {\it et al.}~\cite{Das:2002fr}. The density-dependent elliptic-flow ratios (EFR) in this expression 
are calculated with a modified symmetry potential 
\begin{eqnarray}
V_{\rm sym}(x,\tilde\rho)=\left\{
\begin{array}{l}
V_{\rm sym}^{\rm Gogny}(x,\tilde\rho) \hspace{0.25cm} \text{$\tilde\rho\leq\rho$} \\ 
V_{\rm sym}^{\rm Gogny}(0,\tilde\rho)\hspace{0.25cm} \text{$\tilde\rho>\rho$}
\end{array}
\right.
\eqlab{modsympot}
\end{eqnarray}
with $x = \pm 1$ according to Eq.~(\ref{defrdef}). The difference of the $x = \pm 1$ potentials is tested
only at densities up to the particular $\rho$, the argument of DEFR.
This choice leads to DEFR$^{(n,Y)} (0) = 0$ and to the proper stiff-soft splitting for large values of the 
density $\rho$. Values at intermediate densities are a measure of the impact on elliptic flow observables of 
regions of density lower than that chosen for the argument. 
The derivative of DEFR with respect to density provides thus the sensitivity density 
of the elliptic flow ratio observable under consideration as a function of the nuclear matter density.

\begin{figure}[htb]           
\centering
\includegraphics[width=7.5cm]{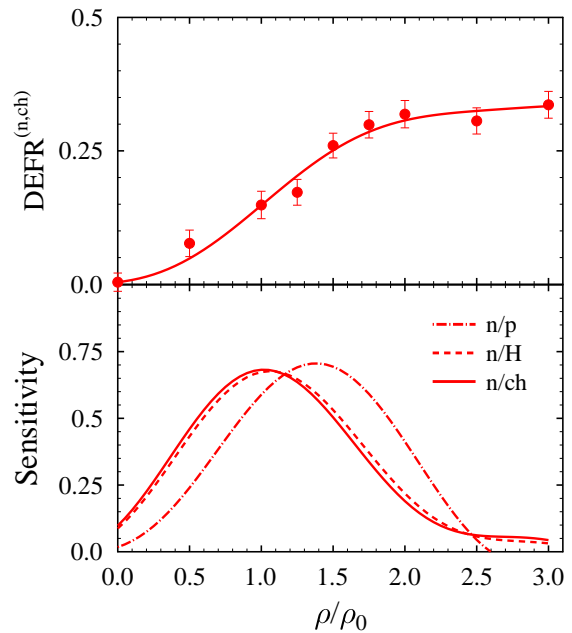}
\caption{Density dependence of the difference of the elliptic flow ratio (DEFR) of neutrons
over charged particles, defined by Eq.~(\protect\ref{defrdef}), for $^{197}$Au+$^{197}$Au collisions 
at 400 MeV/nucleon obtained with the T\"{u}QMD transport model~\protect\cite{Coz13} and 
the FOPI-LAND acceptance filter (top) and the corresponding sensitivity density (bottom, solid line) together with
sensitivity densities obtained from elliptic-flow ratios of neutrons over all hydrogen isotopes (dashed line) and 
neutrons over protons (dash-dotted line).
}
\label{fig:dens}       
\end{figure}

In the top panel of Fig.~\ref{fig:dens}, the density dependence of DEFR$^{(n,Y)}$ for the choice $Y$=all charged 
particles (ch) is presented. It is seen that DEFR increases monotonically up to density values in the neighborhood 
of 2.5\,$\rho_0$, close to the maximum density probed by nucleons in heavy-ion collisions at 400 MeV/nucleon 
incident energy. The relative sensitivity of the elliptic flow ratio of neutrons over charged particles to the various 
density regions is presented in the bottom panel of Fig.~\ref{fig:dens}, together with the same quantity for the 
neutron-over-proton and neutron-over-hydrogen flow ratios.
It is readily observed that the maximum sensitivity of the neutron/proton EFR lies in the 1.4 to 1.5 $\rho_0$ region. 
It is lowered to 1.0 to 1.1 $\rho_0$ for the choices that involve light complex particles. The probed regions of
nuclear density are thus considerably higher than the densities around or below 0.7 $\rho_0$ probed with nuclear 
structure observables~\cite{brown13,zhang13,horowitz13}. Even lower densities in the vicinity of $\rho_0/3$ have
very recently been reported as the region of sensitivity probed with the dipole polarizability 
of $^{208}$Pb~\cite{zhang15}.

The moderately different density regions probed by EFR observables involving protons and, respectively, light complex
particles are expected to lead to slightly different extracted values for the asy-EoS stiffness. Preliminary results, 
employing existing experimental FOPI-LAND data for $v_2^n/v_2^p$ and $v_2^n/v_2^H$ and the T\"{u}QMD transport model, 
suggest this to be the case~\cite{Cozma:2015}.
A slightly stiffer asy-EoS is favored by the latter observable, a difference that will be enhanced if one corrects 
for the fact that 
transport models coupled with phase-space coalescence algorithms tend to underpredict light cluster multiplicities 
by factors ranging up to 2-3. Deuterons and tritons are of particular interest here.
This result suggests that, at higher densities,
the asy-EoS density dependence is somewhat softer than at saturation. It may thus be feasible to extract constraints 
for the parameters of the higher-order terms of the Taylor expansion of the symmetry energy in density around the 
saturation point, in particular the curvature parameter $K_{\rm sym}$. Information regarding the curvature is of high interest 
as, e.g., the inclusion of exchange terms in microscopic models cause a stiffening~\cite{greco04}, while considering the 
momentum tails caused by short-range correlations may cause a softening~\cite{caili16} of the predictions for the 
symmetry energy in the density regime near and above saturation.

It is, therefore, of extreme importance for future experiments to be able to extract
a clean separate proton signal. Additionally, theoretical models that allow for an independent adjustment of the
slope and curvature parameters of the symmetry energy term will be required to be able to push the extracted 
constraints for the asy-EoS
density dependence into the 2$\rho_0$ region.

\section{Conclusion and outlook}
\label{sec:conc}

From the measurement of the elliptic flows of neutrons and light charged particles in the reaction 
$^{197}$Au+$^{197}$Au at 400 MeV/nucleon incident energy a new, more stringent constraint for the nuclear
symmetry energy at suprasaturation density has been deduced. 
From the comparison of the elliptic flow ratio of neutrons over charged particles with UrQMD predictions, 
a value $\gamma = 0.72 \pm 0.19$ is obtained for the power-law coefficient of the potential part in the 
parametrization of the model. 
It confirms the moderately soft to linear density dependence of the symmetry energy deduced previously from the  
FOPI-LAND data. The densities probed were shown to reach beyond twice saturation. 

The effects of deficiencies of the LAND timing electronics have been studied in detail
and their effects assessed by 
systematically varying correction parameters over their intervals of uncertainty. An acceptance-integrated
flow ratio for neutrons over charged particles has been generated by integrating over the time-of-flight spectra.
It is largely insensitive to timing uncertainties but still subject to a systematic error caused by an instrumental
smearing of detection thresholds. Their effect contributes to the total error $\Delta\gamma = \pm 0.19$ of the
acceptance-integrated result that includes a statistical error $\Delta\gamma = \pm 0.10$. 

The slope parameter that corresponds to the obtained parametrization of the symmetry energy is $L = 72 \pm 13$~MeV. 
As densities near and beyond saturation are efficiently
probed with the present observable, one may convert this result into a symmetry pressure 
$p_0 = \rho_0 L/3 = 3.8 \pm 0.7$~MeVfm$^{-3}$ 
(with $\rho_0 = 0.16$~fm$^{-3}$), equivalent to $6.1 \pm 1.1 \times 10^{32}$~Pa. It represents 
the pressure in pure neutron matter at saturation because the pressure in symmetric matter vanishes at this density.
The pressure in neutron-star matter with asymmetries $\delta = (\rho_n - \rho_p)/\rho$ less than unity should be lower. 
The estimate developed in Sec. 9.1 of Ref.~\cite{lipr08}, based on $\beta$ equilibrium, 
yields a proton fraction $x_p = (1-\delta)/2$ of about 5\% for $E_{\rm sym} = 34$~MeV [cf. Eq.~(\ref{eq:pot_term})] 
and saturation density. 
With the corresponding asymmetry $\delta =$~0.90, the pressure of the asymmetric baryonic matter is reduced to 3.1~MeVfm$^{-3}$. 
Adding the contribution of the degenerate electrons yields a value of 3.4~MeVfm$^{-3}$ for the pressure in neutron-star matter 
at saturation density. 
The same or very similar values are obtained with the expressions presented in Refs.~\cite{lattimer01,lattimer14}. 
Compared to the results of Steiner {\it et al.}~\cite{steiner13}, they are located
within the upper half of the 95\% confidence interval obtained by these authors from neutron-star observations. 

While interpretations in this direction may still appear speculative at present and in need of further study, 
they reveal the potential of pressure measurements in nuclear reactions. 
As far as the modeling 
of nuclear reactions is concerned,
it will be important to improve the description of the nuclear interaction in transport models~\cite{junxu16}, 
reduce the parameter ranges also in the isoscalar sector, improve the algorithms used for clusterization,
as well as go beyond the mean-field picture, including short-range correlations. 
The latter have recently been investigated in nuclei~\cite{subedi08,hen14sc} and their consequences 
for transport descriptions of intermediate-energy heavy-ion reactions are of high interest and need to be investigated~\cite{hen15}.
Moreover, it will be quite important to compare the experimental data with the predictions of several transport models, 
of both Boltzmann-Vlasov and molecular-dynamics type~\cite{guoyongwang13}, to pursue the work towards a 
model-independent constraint of the high-density symmetry energy initiated in Ref.~\cite{Coz13}.
 
The results of the present experiment, together with the theoretical study of the density probed, 
may also be seen as a strong encouragement for 
extending the measurement of neutron and charged particle flows to other reaction systems and energies. 
The presented calculations suggest that the curvature parameter $K_{\rm sym}$ can be addressed experimentally
if higher precision and elemental and isotopic resolution for charged particles can be achieved.  
Future experiments will, therefore, benefit from the improved calorimetric capabilities of the NeuLAND detector 
presently constructed as part of the $R^{3}B$ experimental 
setup~\cite{NeuLAND} and from the availability 
of radioactive ion beams for reaction studies at FAIR. 

\begin{acknowledgments}
The authors are indebted to 
the Accelerator Department and the Target Laboratory of the GSI Helmholtzzentrum for providing 
high-quality beams and targets. We are particularly grateful to 
the Laboratori Nazionali del Sud for making parts of the CHIMERA multidetector available for the experiment.
We thank W.~Reisdorf and the FOPI Collaboration for providing specifically selected data sets from their experiments 
and for continuing support of the project.
The contributions of R.~Bassini and C.~Boiano during the preparatory phase are gratefully acknowledged.

This work has been supported by the European Union under Contract No. FP7-25431 (Hadron-Physics2), 
by INFN (Istituto Nazionale di Fisica Nucleare)
experiments EXOCHIM and NEWCHIM,
by the National Natural Science Foundation of China under Grants No. 11375062, No. 11547312, and No. 11505057, 
by the Hungarian OTKA Foundation No. K106035, by the Polish Ministry of Science and Higher Education under Grant No. DPN/N108/GSI/2009, 
by the Foundation for Polish Science - MPD program, co-financed by the European Union within the European Regional Development Fund, 
by the Polish National Science Center (NCN), Contracts No. UMO-2013/10/M/ST2/00624 and No. UMO-2013/09/B/ST2/04064,
by the Slovak Scientific Grant Agency under Contract 2/0121/14,
by the UK Science and Technology Facilities Council (STFC) under Grants No. ST/G008833/1, No. ST/I003398/1, and No. STBA00019,
by the U.S. Department of Energy under Grants No. DE-FG02-93ER40773  and No. DE-SC0004835, by the U.S. National Science Foundation 
Grant No. PHY-1102511, and by the Robert A. Welch Foundation through Grant No. A-1266. 
\end{acknowledgments}

\section{Appendix: Correction of LAND timing}
\label{sec:app}

In the TACQUILA electronic board~\cite{Koch05},  
the time measurement of a recorded hit is performed by registering the 
time of the start signal (tac) inside a 25-ns clock-cycle window, 
the time of the common stop signal inside its
25-ns clock-cycle window (so-called $t_{17}$), and the number $n_c$ of cycles occurring between the start 
and stop cycles. The returned calibrated time information $t_{\rm cal}$ is then given by  
\begin{equation}
t_{\rm cal}={\rm tac}+25n_c-t_{17}~({\rm ns}).
\end{equation}
The resolution of the timing system is of the order of 10~ps (rms)~\cite{Koch05}.

\begin{figure}[!t]              
\centering
\includegraphics[width=7.5cm]{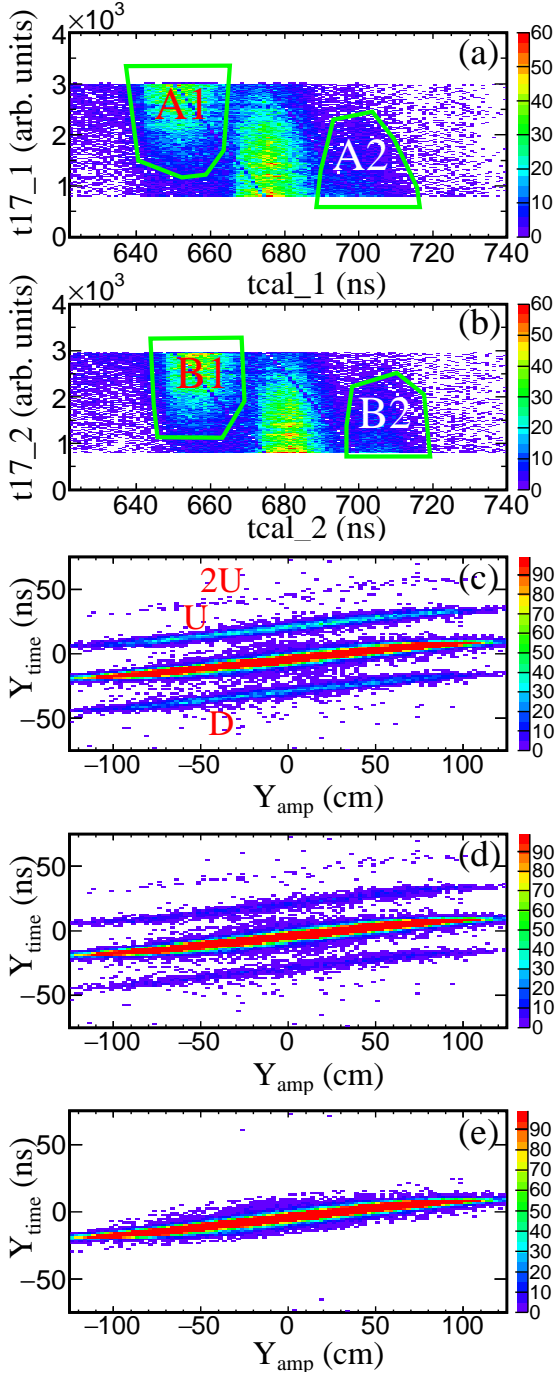}
\caption{(a),(b) Observed correlations of $t_{17}$ vs calibrated time $t_{\rm cal}$ 
for the two signals $t_{\rm cal\_1}$ and $t_{\rm cal\_2}$ of a module of the first plane of LAND, respectively.
(c),(d) Observed correlations of the position 
signals $Y_{\rm time}$ vs $Y_{\rm amp}$ deduced from the time and amplitude information of these signals, 
respectively, before (c) and after (d) the first correction step. (e) the same correlation after the  
correction step $1^{st}$bis. The significance of the marked regions in panels (a) through (c) is explained 
in the text.
}
\label{fig:corra}       
\end{figure}

Examples of the observed correlations between $t_{17}$ (in channels) and the so measured $t_{\rm cal}$ (in~ns) 
for the two photomultipliers (PM's) of a paddle of the first plane of LAND are shown in panels (a) and (b) of 
Fig.~\ref{fig:corra}, respectively. Ideally, no correlation should be visible as the distribution of the stop signals 
inside the clock cycle window should be completely random. 
Unexpectedly, however, a strong correlation is observed; preferences exist, primarily, for 
high $t_{17}$ values at smaller times $t_{\rm cal}$ but also for low $t_{17}$ values at larger times. 
This behavior by itself implies an improper functioning of the TACQUILA board. 
It is evidence of incorrect determinations of $n_c$, depending on where 
the $t_{17}$ signal appears within the clock cycle. 
In addition, it was found that the probability of wrong $n_c$ countings was rate dependent; it increased 
with increasing frequency of hits recorded in the LAND modules. This behavior, as discovered during the
data taking was confirmed with bench tests performed after the experiment and ultimately 
corrected by replacing parts of the TACQUILA electronic readout system. 
 
As a consequence, the region marked as A1 in 
Fig.~\ref{fig:corra}~(a) must be considered as
overpopulated because of a wrong counting of the number $n_c$ of clock cycles; 
the returned $n_c$ is likely to be one unit 
smaller than the true value, causing an offset of -25~ns of the calibrated time $t_{\rm cal}$. 
With smaller probability, counting errors larger than one cycle were observed as well. 
It follows that any measured time in LAND is not necessarily but possibly wrong by 
$\pm 25$~ns or, with decreasing probability, even multiples of it.  

The described malfunctions clearly affect the measurements of the hit position $Y_{\rm time}$ along the 
vertically oriented paddles, derived from 
the difference, and of the arrival time $t_{\rm hit}$ at the paddle, derived from the sum of the two signals recorded 
for a hit. The two quantities 
are given by  
\begin{equation}Y_{\rm time}=t_{\rm cal\_1}-t_{\rm cal\_2}\end{equation}
\begin{equation}t_{\rm hit}=(t_{\rm cal\_1}+t_{\rm cal\_2})/2\end{equation}
where the indices 1 and 2 refer to the two PM's of a given paddle; 
the signals $t_{\rm cal\_1}$ and $t_{\rm cal\_2}$ are, at this stage, not yet synchronized, i.e. not yet
corrected for time offsets generated by, e.g., differences of the cable lengths of the two PMs. The position
$Y_{\rm time}$ is, therefore, still given in units of nanoseconds and not necessarily centered with respect 
to the paddle length.

In the case of malfunctions, the time differences may be sufficiently large, so that the deduced hit position falls
outside the physical length of the paddle. This can be easily corrected by adding or subtracting 25~ns 
to the time difference. It will move the hit to its correct position inside the paddle. To recover 
the correct arrival times $t_{\rm hit}$ is not equally feasible in this case. 
It would require the knowledge of whether the wrong position of $Y_{\rm time}$ 
is caused by erroneous +25~ns in one or -25~ns in the other of the two signals coming from a paddle. 
The two possibilities correspond to $t_{\rm hit}$ values that differ by 25~ns. Moreover, it is also possible 
that both time measurements are affected by the same $\pm 25$~ns time jump. In that case, the position 
$Y_{\rm time}$ is correct but the returned arrival time $t_{\rm hit}$ is erroneous by $\pm 25$~ns. 
Because the expected range of arrival times at LAND exceeds 25~ns, an easy and straightforward procedure for recovering 
the correct time information does not exist.

It has, nevertheless, been possible to develop a correction scheme for recovering the correct 
times with high probability and for determining the consequences of remaining uncertainties 
for the finally determined symmetry-term coefficient. This was achieved with the help of correction parameters 
whose effects can be assessed on a quantitative level. The scheme divides into two parts.

\begin{figure}[htb]            
\centering
\includegraphics[width=8.5cm]{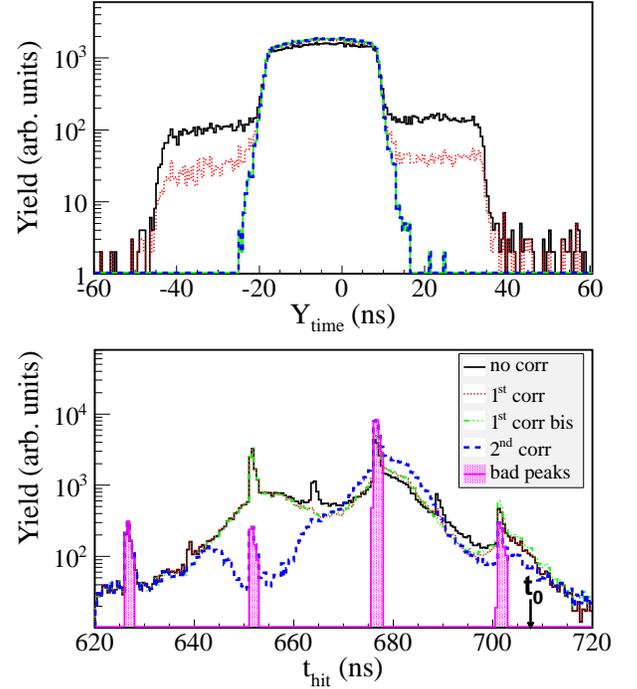}
\caption{Hit distributions for a module of the first plane of LAND
as a function of the position $Y_{\rm time}$ (top) and of the arrival time $t_{\rm hit}$ (bottom). 
Black solid lines denote the uncorrected 
distributions, the colored lines represent the distributions after the $1^{st}$ (red, small dots), 
the $1^{st}$bis (green, thick dots), and the $2^{nd}$ (blue, thick dashed)
corrections. ``Bad peaks'' refers to artificial sharp peaks at 25~ns intervals (solid purple areas)
generated by the electronic readout.
}
\label{fig:corrb}       
\end{figure}

The first correction step starts from the observed correlation of the position measurement $Y_{\rm time}$ 
with the position $Y_{\rm amp}$ obtained from the amplitudes of the normalized PM signals. The uncorrected
correlation [Fig.~\ref{fig:corra}(c)] shows clearly separated regions of unphysical positions
$Y_{\rm time}$, marked with U (up), D (down), and 2U (twice up), in addition to the strongest group of 
coinciding position measurements.
The distribution of uncorrected positions $Y_{\rm time}$ for a typical module of the first plane of LAND
is shown in the top panel of Fig.~\ref{fig:corrb} and the corresponding 
$t_{\rm hit}$ distribution is shown in the bottom panel of the same figure (``no corr,'' solid line in black). 
The two side groups with wrong $Y_{\rm time}$ positions are weak ($<10\%$) compared to the main group 
but significant. The probability for double time jumps in the same direction is below 1\% and essentially 
negligible. The 25-ns repetitions of structures in $t_{\rm hit}$ are 
clearly visible in Fig.~\ref{fig:corrb} (bottom panel), in particular, the repeated appearance of narrow 
artificial peaks generated by the electronics. These structures were removed before other corrections were applied.

In the attempt to correct the wrong positions, the chosen scheme takes into account the value of $t_{17}$
relative to the returned time as shown in panels (a) and (b) of Fig.~\ref{fig:corra}. 
In the example of a hit belonging to the region 'U' in panel (c), $t_{\rm cal\_1}$ may be located in what is defined
as the ``good'' region, i.e., in the interval between 640~ns and 720~ns but outside the gates 'A1' and 'A2' 
in panel (a), and $t_{\rm cal\_2}$ may be located in region 'B1' 
of panel (b). In this case, it is obviously more probable that $t_{\rm cal\_2}$ is incorrect, i.e., that the number
of clock cycles is wrong by one unit, and 25~ns are thus added to its value. Instead, if $t_{\rm cal\_1}$ is located 
in region 'A2' and $t_{\rm cal\_2}$ in the ``good'' region, i.e, outside the gates 'B1' and 'B2' in panel (b), 
25~ns are subtracted from $t_{\rm cal\_1}$. Corresponding corrections are applied to hits
belonging to regions marked as 'D' and '2U' in Fig.~\ref{fig:corra}(c), as well as to a region '2D' when it appeared 
in other cases. This part of the 
correction scheme is summarized in Table~\ref{tab:corr} and marked as $1^{st}$ step in the last column. Note that three
possibilities exist for correcting the rare double jumps, depending on where the hits are found to be located. The superscripts
``+'' and ``-'' used in the table indicate regions analogous to the four regions 'A1', 'A2', 'B1', 
and 'B2' marked in panels (a) and (b) of Fig.~\ref{fig:corra} but located further out by another +25~ns or -25~ns from the central 
part of the spectrum. 
 
\begin{table}[ht]
\begin{tabular}{|c|c|c|r|r|c|}
\hline
 panel (c) &  panel (a) &  panel (b) & $t_{\rm cal\_1}$ & $t_{\rm cal\_2}$ & correction \\ \hline
U           & A2          &             & -25 ns     &            & $1^{st}$ \\ \hline
U           &             & B1          &            & +25 ns     & $1^{st}$ \\ \hline
D           & A1          &             & +25 ns     &            & $1^{st}$ \\ \hline
D           &             & B2          &            & -25 ns     & $1^{st}$ \\ \hline
2U          & A2          & B1          & -25 ns     & +25 ns     & $1^{st}$ \\ \hline
2U          & ~~A2$^{+}$  &             & -50 ns     &            & $1^{st}$ \\ \hline
2U          &             & ~~B1$^{-}$  &            & +50 ns     & $1^{st}$ \\ \hline
2D          & A1          & B2          & +25 ns     & -25 ns     & $1^{st}$ \\ \hline
2D          & ~~A1$^{-}$  &             & +50 ns     &            & $1^{st}$ \\ \hline
2D          &             & ~~B2$^{+}$  &            & -50 ns     & $1^{st}$ \\ \hline
good        & A1          & B1          & +25 ns     & +25 ns     & $2^{nd}$ \\ \hline
good        & A2          & B2          & -25 ns     & -25 ns     & $2^{nd}$ \\ \hline
\end{tabular}
\caption{The first three columns indicate the regions referred to in the listed panels of Fig.~\protect\ref{fig:corra} 
while the next two columns specify the actions taken on $t_{\rm cal\_1}$, $t_{\rm cal\_2}$, or both. The
last column indicates the number of the correction step as given in the text.
}
\label{tab:corr}
\end{table}

Panel (d) of Fig.~\ref{fig:corra} shows the $Y_{\rm time}$-vs-$Y_{\rm amp}$ correlation after this
first correction step. The corresponding $Y_{\rm time}$ and $t_{\rm hit}$ distributions are shown in 
Fig.~\ref{fig:corrb} (in red).
It is evident that not all the wrong positions have disappeared because some hits do not fulfill the 
assumptions made in devising the first step of the correction scheme (of the order of 2\%, cf.
Fig.~\ref{fig:corrb}, top panel). In that case, an additional
correction called $1^{st}$bis is applied. At this step, the location of the hit pattern in the 
$t_{17}$-vs-$t_{\rm cal}$ maps [Figs.~\ref{fig:corra}(a) and \ref{fig:corra}(b)] is ignored and the 
correct $Y_{\rm time}$ is recovered by either adding 25~ns to one or by subtracting 25~ns from the other
of the two time signals $t_{\rm cal\_1}$ and $t_{\rm cal\_2}$ of that hit. The choice made between these
two possibilities was based upon which of them had appeared with the higher probability when the $1^{st}$
correction step had been applied to the same paddle. 
Panel~(e) of Fig.~\ref{fig:corra} shows the $Y_{\rm time}$-vs-$Y_{\rm amp}$ 
correlation after this correction: Now all the positions deduced from time signals are correct. They
coincide with the positions deduced from  the amplitudes and are within the physical length of the
paddle (Fig.~\ref{fig:corrb}, top, in green, coinciding with blue).

At this stage, cases in which both time measurements are affected by the same time jump have not been 
touched. They remain correct regarding their positions $Y_{\rm time}$ but the problem of their erroneous 
arrival times $t_{\rm hit}$ is not solved yet. 
For that purpose, an additional correction step has been conceived. It is based on the assumption that
the coincident location of the two signals of a hit in either regions 'A1' and 'B1' or in 'A2' and 'B2' 
of their respective $t_{17}$-vs-$t_{\rm cal}$ maps is a strong indication of a simultaneous jump. 
The correction step consists of either adding or subtracting 25~ns to both values 
$t_{\rm cal\_1}$ and $t_{\rm cal\_2}$ of that hit, so that they fall into the central regions 
of their maps. It is marked as $2^{nd}$ step in the last column of Table~\ref{tab:corr}. It simply 
changes the arrival times $t_{\rm hit}$ by 25~ns but leaves the position $Y_{\rm time}$ and its
correlation with $Y_{\rm amp}$ unaffected.  
The so-obtained final $t_{\rm hit}$ distribution is shown in 
Fig.~\ref{fig:corrb} (bottom panel, in blue).

\begin{figure}[!t]            
\includegraphics[width=8cm]{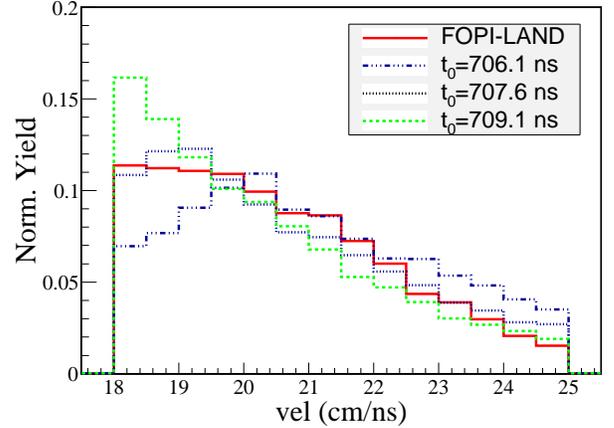}
\caption{High velocity tail of normalized velocity spectra for several assumptions on the 
time-zero value $t_0$ in comparison with the corresponding spectrum measured in the FOPI-LAND experiment. }
\label{fig:tzero} 
\end{figure}

It is evident that the second correction step falsely modifies correctly measured cases of long or 
short times with time signals $t_{\rm cal\_1}$ and $t_{\rm cal\_2}$ accidentally falling into the marked regions.
Its effect is particularly large in the interval 640 to 660~ns of the $t_{\rm hit}$ spectrum, where it
causes a depression (Fig.~\ref{fig:corrb}, bottom panel).  
The geometric mean between the yields before and after this correction would approximately represent
a smooth time spectrum that would seem more probable. This level can be reached if only about 80\% of the 
hits near the center down to about 50\% towards the edges of this region are actually moved in
the second step, while the rest of the selected candidates are left at their original arrival times in the
640~ns-to-660~ns interval. However, as it is not known which of the hits should be moved and which  
should be left at their time positions, a correction of this kind is not properly feasible. 
It will smoothen the time spectrum but,
because of the necessarily random selection, an inevitable mixing of hits between the affected time intervals 
will occur.

This situation was addressed by considering the fraction
of randomly selected hits whose arrival times are actually modified in step 2 as an unknown correction parameter. 
The time spectrum in Fig.~\ref{fig:corrb} (bottom panel) and the comparison of flow results as a function of this 
fraction with FOPI results (Fig.~\ref{fig:fopiflow} in Sec.~\ref{sec:timcorr}) suggest a value of at least 40\%. 
Apart from that, it remains unknown and its significance for the differential flow ratios must be assessed. 
The result, a systematic variation of $\Delta\gamma=0.05$ as a function of this fraction,
is shown in Fig.~\ref{fig:fraction} and discussed in Sec.~\ref{sec:diffdat}. For the acceptance-integrated
analysis based on time-integrated data sets (Sec.~\ref{sec:integ}), the present corrections are of minor importance
because very few hits are actually moved across the boundaries of the integration interval. 

Owing to the logarithmic gain chosen for the new TACQUILA electronic board, the signals of low-energy $\gamma$ 
rays fell below threshold with the effect that the calibration of the time spectra could not be based on
a measured $\gamma$ peak. The location of the zero-time-of-flight point $t_0$ was, therefore, determined
from a comparison of velocity spectra, generated with various assumptions on $t_0$, with the well-calibrated
spectrum available from the FOPI-LAND experiment. 
The high-velocity part of the spectrum was found to exhibit the most distinctive variation as a function of 
the choice for $t_0$ (Fig.~\ref{fig:tzero}). The presence of artificial peaks at arrival times 
$t_{\rm hit} \approx 677$ and 702~ns (Fig.~\ref{fig:corrb}, bottom panel)  
limited the useful range to velocities $vel > 18$~cm/ns or $E_{\rm kin} > 230$~MeV for nucleons. The rapid 
variation of the velocity spectrum with the choice of $t_0$ permitted its determination with an uncertainty 
of the order of 1~ns (Fig.~\ref{fig:tzero}). Its location at $t_{\rm hit} = 707.6$~ns is marked in the 
spectrum of arrival times $t_{\rm hit}$.  
As the displayed times are measured with respect to a delayed common stop signal, finite time-of-flight
values are to the left of $t_0$. Photons would appear at $t_{\rm hit} = 691$~ns, indicating that the yield at 
larger $t_{\rm hit}$ represents the level of background and of hits that are still misplaced. 
The interval $18 \le vel < 25$~cm/ns used for the
comparison corresponds to $680 \le t_{\rm hit} < 688$~ns, a region only mildly affected by corrections. 
The same is true for the main group of recorded hits with arrival times between $t_{\rm hit} = 669$ and 685~ns, 
corresponding to flight times between 23 and 39~ns and to kinetic energies of 100 to 400 MeV for nucleons 
(note that artificial peaks are removed).
 
The correction effects are stronger for arrival times between $t_{\rm hit} = 642$ and 663~ns, expected for nucleons 
with approximately 30 to 70 MeV kinetic energy. The time spectrum in that region is strongly modified by the second 
correction step moving particles from this region into the main group centered at 
$t_{\rm hit} = 680$~ns (Fig.~\ref{fig:corrb}, bottom panel). 
The threshold energy of 60 MeV for protons to pass 
through the veto wall and to be detected in a LAND module is located within the affected 
region ($t_{\rm hit} = 659$~ns). The same is true for the thresholds of deuterons and tritons, located 
at smaller energy per nucleon and correspondingly longer times of flight. 

To be independent of the applied corrections, the acceptance-integrated result was obtained by integrating
the time spectra up to $t_{\rm hit} = 640$~ns, i.e., beyond the critical regions. The maximum time of flight of
67.5~ns defines a threshold of 30~MeV for neutrons. It is lower than the physical thresholds for charged 
particles, a condition that was equally applied in the UrQMD simulations. Only double time jumps and 
background events, apart from the neutrons below threshold, can contribute to the low-intensity region at 
$t_{\rm hit} < 640$~ns. Possible systematic effects related to these effects were investigated by varying the 
integration limit between $617 < t_{\rm hit} < 648$~ns, i.e., flight times between 60 and 90~ns, and by 
correspondingly adjusting the neutron-energy threshold in the calculations. The resulting variation 
of $\gamma$ is small as shown in Fig.~\ref{fig:tofinteg}.

It is once more noted here that the timing corrections are applied to all particles independently of whether 
they are charged or neutral. This has obviously reduced their influence on the flow ratios that are
used as the principal observables, in agreement with the results of the tests performed.

\end{document}